\documentclass[final,5p,times,twocolumn]{elsarticle}
\usepackage{amssymb}
\usepackage{amsmath}
\usepackage[cmyk,dvipsnames]{xcolor}
\usepackage{placeins}
\usepackage[utf8]{inputenc} 
\usepackage[T1]{fontenc}    
\usepackage{booktabs}       
\usepackage{amsfonts}       
\usepackage{nicefrac}       
\usepackage{microtype}      
\usepackage{lipsum}
\usepackage{graphicx}
\graphicspath{ {./images/} }
\usepackage{hyperref}
\usepackage{titletoc}
\hypersetup{
    colorlinks,
    linkcolor={blue},
    citecolor={blue},
    urlcolor={blue}
}

\renewcommand{\vec}[1]{\mathbf{#1}}

\usepackage[resetlabels]{multibib}
\newcites{supple}{References}

\makeatletter
\def\ps@pprintTitle{%
  \let\@oddhead\@empty
  \let\@evenhead\@empty
  \let\@oddfoot\@empty
  \let\@evenfoot\@oddfoot
}
\makeatother

\begin{document}

\begin{frontmatter}

    \title{Flowy: High performance probabilistic lava emplacement prediction}

    \address[UI]{Science Institute and Faculty of Physical Sciences, University of Iceland VR-III,107 Reykjav\'{\i}k, Iceland}
    \address[RWTH]{Department of Physics, RWTH Aachen University, 52056 Aachen, Germany}
    
    \author[UI,RWTH]{Moritz Sallermann} 
    \author[UI]{Amrita Goswami}
    \author[UI]{Alejandro Peña-Torres}
    \author[UI]{Rohit Goswami}
    
    \begin{abstract}
        Lava emplacement is a complex physical phenomenon, affected by several factors.
        These include, but are not limited to features of the terrain, the lava settling process, the effusion rate or total erupted volume, and the probability of effusion from different locations. One method, which has been successfully employed to predict lava flow emplacement and forecast the inundated area and final lava thickness, is the MrLavaLoba method from Vitturi \textit{et al.}~\cite{demichielivitturiMrLavaLobaNewProbabilistic2018}. The MrLavaLoba method has been implemented in their code of the same name~\cite{lava_loba_github}.
        Here, we introduce Flowy, a new computational tool that implements the MrLavaLoba method in a more efficient manner. New fast algorithms have been incorporated for all performance critical code paths, resulting in a complete overhaul of the implementation.
        When compared to the MrLavaLoba code~\cite{demichielivitturiMrLavaLobaNewProbabilistic2018,lava_loba_github}, Flowy exhibits a significant reduction in runtime  -- between 100 to 400 times faster -- depending on the specific input parameters.
        The accuracy and the probabilistic convergence of the model outputs are not compromised, maintaining high fidelity in generating possible lava flow paths and deposition characteristics.
        We have validated Flowy's performance and reliability through comprehensive unit-testing and a real-world eruption scenario.
        The source code is freely available on GitHub~\cite{flowy_github}, facilitating transparency, reproducibility and collaboration within the geoscientific community.
        \end{abstract}
        
        
        \end{frontmatter}
        
        \section{Introduction}
        Understanding the different types of lava flows is essential for anticipating potential disasters during volcanic eruptions around the world.
        However, the complex behavior of these flows makes accurate predictions of their paths challenging.
        To tackle this issue, several numerical models and simulation tools have been developed over the past decades.
        These existing lava flow models vary significantly in terms of complexity, underlying assumptions, dimensionality and accessibility. 
        
        Typically, these models are divided into deterministic and probabilistic (or stochastic) ones.
        The deterministic models try to replicate the dynamic behavior of natural systems with varying levels of fidelity.
        By doing so, they predict the emplacement of erupted volumes over time, provided that the input parameters are correctly set.
        Examples of codes implementing such models include  MULTIFLOW~\cite{richardson2019multi}, VolcFlow~\cite{kelfoun2016volcflow}, MAGFLOW~\cite{vicari2007modeling}, its successor GPUFLOW~\cite{cappelloModelingGeophysicalFlows2022}, FLOWGO~\cite{harrisFLOWGOKinematicThermorheological2001} and its re-implementation PyFLOWGO~\cite{harrisSimulatingThermorheologicalEvolution2016,chevrelPyFLOWGOOpensourcePlatform2018}.
        A comprehensive review and benchmark of the current state of deterministic models can be found in Refs.~\cite{cordonnier2016benchmarking,dietterichBenchmarkingComputationalFluid2017}.
        On the other hand, probabilistic models are based on the fact that lava can be considered to be a gravitational current that follows the path of steepest descent based on the topography of the terrain, with random deviations.
        In this work, we focus on the probabilistic approach to model lava flows.
        
        Amongst the stochastic models that have been developed and used to simulate lava flows are DOWNFLOW~\cite{favalliForecastingLavaFlow2005}, Q-LAVHA~\cite{mossouxQLAVHAFlexibleGIS2016} and MrLavaLoba~\cite{demichielivitturiMrLavaLobaNewProbabilistic2018}.
        The MrLavaLoba method and code have been extensively, and successfully, used in the recent years to perform lava hazard assessments in Iceland~\cite{tarquiniModelingLavaFlow2018,tarquiniAssessingImpactLava2020,barsottiEruptionFagradalsfjall20212023,pedersenLavaFlowHazard2023}.
        Two key advantages of the MrLavaLoba method are its ability to account for lava flow volume and modifications to the terrain topography caused by the lava itself.
        This is especially valuable to assess the efficacy of barriers in diverting lava flows.
        Some examples, which received worldwide attention, are the protective walls, that have been erected to shield the city of Grindav\'{i}k from the lava emitted in the 2024 eruption~\cite{UnrestGrindavik}.
        
        Nevertheless, the MrLavaLoba \emph{code} has several limitations.
        The runtime for simulations can become rather long in large scenarios, which limits the ability to respond in real-time, during on-going volcanic eruptions when the conditions and terrain can change rapidly.
        Additionally, this reduces the statistical quality of the results since obtaining enough samples for statistical measures to hold is frequently impractical.
        The long runtime can also make enhancing the resolution of the terrain mesh infeasible. Therefore, downsampling of the digital elevation model (of the terrain) is necessary.
        Moreover, creating large-scale hazard maps -- crucial for hazard assessment and planning risk mitigation measures --  can require hundreds of thousands of simulations.
        Both simulation runtime, and data storage, can become prohibitive in such situations.
        
        In this paper, we present Flowy, a new high-performance probabilistic code to forecast lava flow inundation efficiently.
        Flowy implements the MrLavaLoba \emph{method} of lava emplacement established by Vitturi \textit{et al.}~\cite{demichielivitturiMrLavaLobaNewProbabilistic2018}, which itself is based on the foundational work of Glaze \textit{et al.}~\cite{glazeSimulationInflatedPahoehoe2013}.
        Flowy significantly enhances the simulation efficiency by replacing core algorithms with more performant alternatives.
        Further, we show that Flowy can reduce data storage loads by saving binary output files in NetCDF format~\cite{rewNetCDFInterfaceScientific1990} that can be reduced in size by up to one to two orders of magnitude, via a variety of effective size reduction strategies.
        
        The remainder of this work is organized as follows.
        In Section~\ref{sec:model} we summarize the core of the MrLavaLoba method of Vitturi \textit{et. al.}~\cite{demichielivitturiMrLavaLobaNewProbabilistic2018}, while the details of the improved algorithms are explained in Section~\ref{sec:method}. Next, an overview of the Flowy code is presented in Section~\ref{sec:code_overview}.
        In Section~\ref{sec:results} we assess and discuss the performance of Flowy using a parabolic toy topography and a case study of the Fagradalsfjall eruption of 2021, with parameters derived from Pedersen \textit{et al.} ~\cite{pedersenLavaFlowHazard2023}. 
        Finally, we conclude in Section~\ref{sec:conclusions}.
        
        \section{Method}\label{sec:model}
        \begin{figure}[t]
            \centering
            \includegraphics[width=\linewidth]{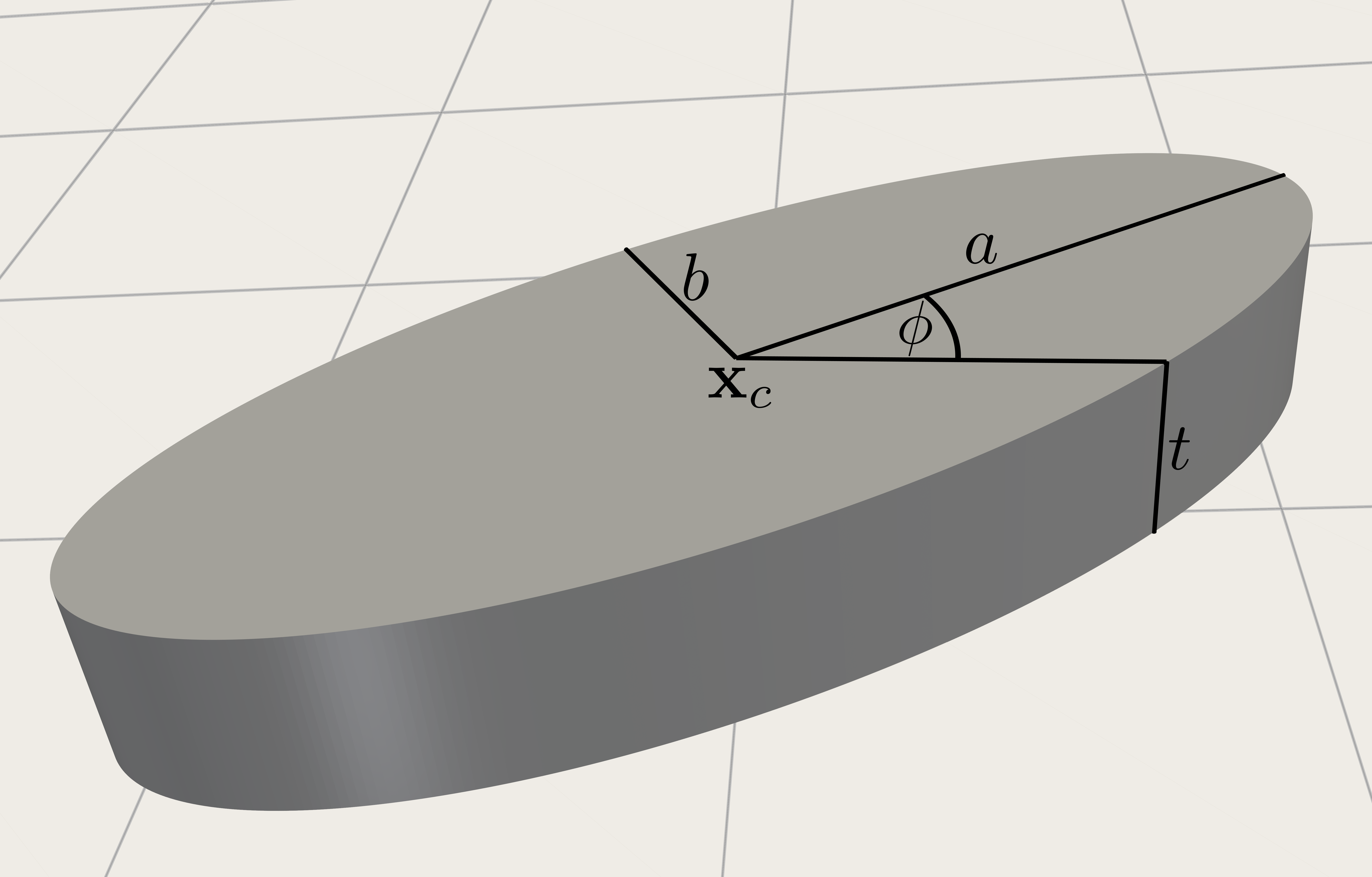}
            \caption{Schematic of an elliptical lobe, which is the fundamental building block of the MrLavaLoba method of Vitturi \textit{et. al.}~\cite{demichielivitturiMrLavaLobaNewProbabilistic2018}. 
            The coordinates of the center, $\vec{x}_c$, the major and minor semi-axes $a$ and $b$, the azimuthal angle $\phi$, and the thickness of the lobe $t$ are marked. The grid lines illustrate the cells of the Digital Elevation Model (DEM).}
        \label{fig:lobe}
        \end{figure}
        
        \begin{figure*}[t]
            \centering
            \includegraphics[width=\linewidth]{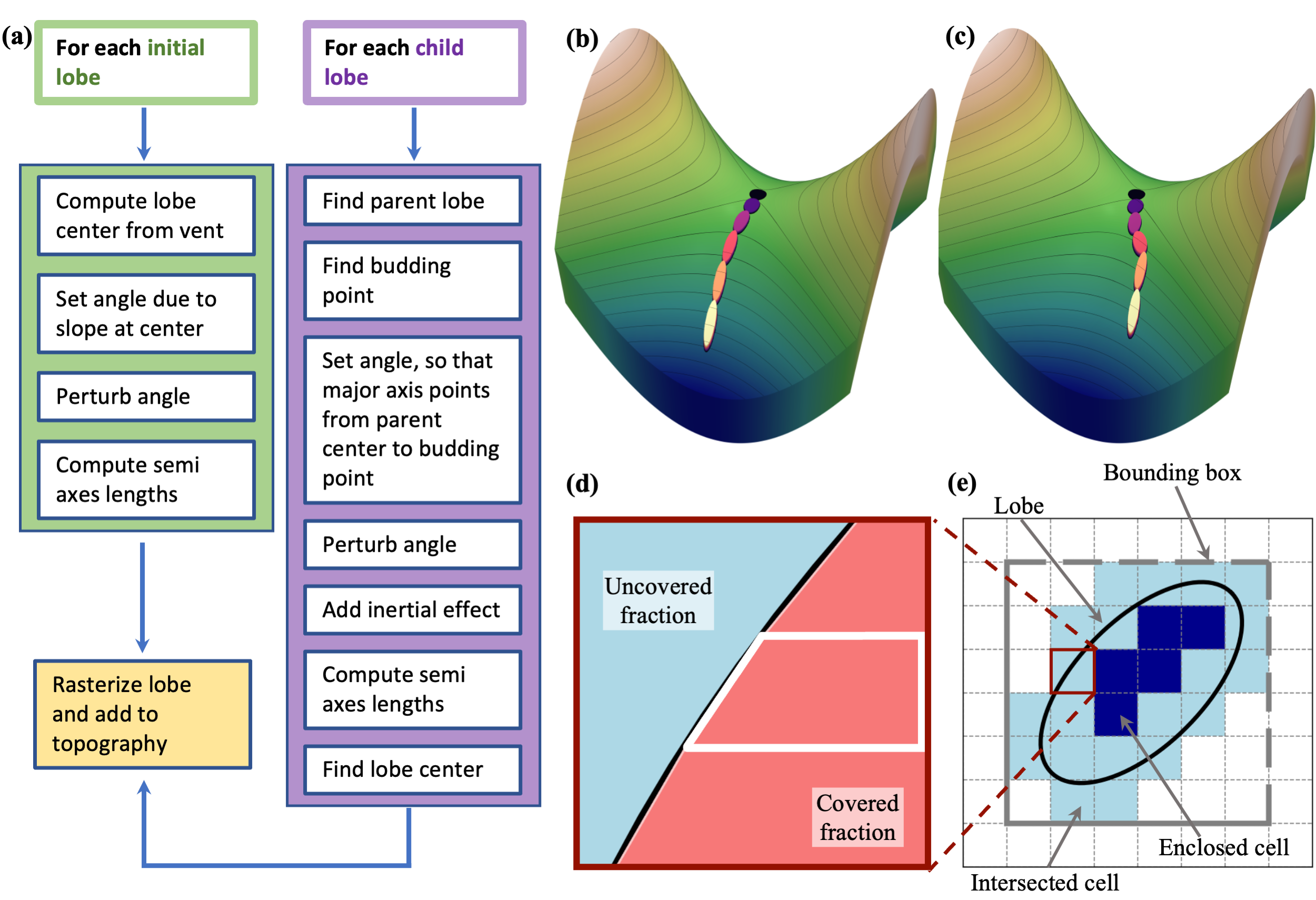}
            \caption{\textbf{(a)} Flow-chart of the MrLavaLoba method of Vitturi \textit{et al.}~\cite{demichielivitturiMrLavaLobaNewProbabilistic2018}, as implemented in Flowy.
            \textbf{(b)} Snapshot of simulation output with six lobes deposited on a saddle topography. The vent is on the saddle point.
            The initial lobe (black) is nearly spherical, because the topography is flat, and later lobes are elongated, due to the increasing slope.
            The thickness of the lobes increases monotonically from the initial to the final.
            \textbf{(c)} Lobe propagation on a saddle topography, such that randomness is added and the lobes deviate from the steepest descent direction.
            \textbf{(d)-(e)} Schematic of a lobe (perimeter depicted as a solid black curve).
            Enclosed (intersected) cells are shaded in dark-blue (light-blue). Solid and dashed grey lines denote the bounding box of the ellipse.
            The inset (d) visually depicts the calculation of the area fraction of a particular cell, outlined in maroon in part e).
            The area fraction is determined using the trapezoidal quadrature rule; scanning line segments are white.
            The uncovered fraction (covered fraction) is light-blue (dark-peach).
            A single trapezoid is showcased (white border), illustrating the trapezoidal quadrature rule used to calculate the covered fraction.}
        \label{fig:cells_rasterized}
        \end{figure*}
        
        In the following, we briefly summarize the essentials of the MrLavaLoba lava emplacement method~\cite{demichielivitturiMrLavaLobaNewProbabilistic2018}.  
        Simulations act on an input topography, which is represented by a digital elevation model (DEM).
        This is in the form of a two-dimensional square grid, with a finite cell size (typically a few meters), wherein every cell encodes the local elevation (see also \ref{sec:grid}).
        
        The overarching structure of a simulation is as follows: 
        \begin{enumerate}\label{sec:sim_structure}
            \item Every simulation, or run, consists of a certain number of flows. 
            \item Every flow is a chain of lobes. Flows are added to the topology sequentially, and are consequently, inter-dependent. Figure~\ref{fig:cells_rasterized}(b) and Figure~\ref{fig:cells_rasterized}(c) show how lobes are deposited on a saddle topography in a single flow.
            \item Each lobe is also successively deposited on the topography, and therefore, lobes are dependent on each other.
        \end{enumerate} 
        
        As illustrated in Figure~\ref{fig:lobe}, a lobe is an ellipse with finite thickness, defined by
        \begin{itemize}
            \item the coordinates of the center $\vec{x}_c$,
            \item the size of the major and minor semi-axes $a$ and $b$,
            \item the orientation of the semi-major axis relative to the Cartesian axes, given by the azimuthal angle $\phi$
            \item and the aforementioned thickness $t$.
        \end{itemize}
        
        These elliptical lobes are progressively deposited on the topography, modifying it during the course of the simulation. 
        The placement and shape of each lobe is influenced by the local slope -- lobes tend to, on average, follow paths of steepest descent, becoming more elongated at steeper terrain gradients. For an example refer to Figure~\ref{fig:cells_rasterized}(b) and Figure~\ref{fig:cells_rasterized}(c).
        
        The following sections describe the generation and placement of lobes in a particular flow.
        They are summarized in the flowchart of Figure~\ref{fig:cells_rasterized}(a). 
        
        \subsection{Initial lobe generation}
        "Initial" lobes  originate from a vent, or from user-defined regions of the topography.
        The location of these vents is an input parameter.
        \begin{itemize}
            \item The lobe center, $\vec{x}_c$, is determined, depending on the supplied vent locations.
            \item The slope of the topography at $\vec{x}_c$ is estimated via bi-linear interpolation of the elevation data.
            The elevation data forms a regular grid.
            \item The azimuthal angle, $\phi$, is set such that the semi-major axis of the lobe points into the direction of steepest descent.
            \item \label{step:angle_perturbation} A perturbation of $\phi$ can optionally add some randomness to the lobe semi-major axis direction. This perturbation in the angle is drawn from a truncated normal distribution on the interval $[-\pi,\pi]$, with zero mean and a standard deviation that is inversely proportional to the magnitude of the slope. In other words, this means that the angle perturbation is smaller for steeper slopes. For more details, refer to Section S2 of the Supplemental Material.  
            \item The lobe semi-axes, $a$ and $b$ can be calculated from the previously estimated slope.
            The aspect ratio $\frac{a}{b}$ linearly increases with the magnitude of the slope (up to a user-defined maximum).
            \item \label{step:thickness} Finally, the thickness, $t$, is determined, according to a user-defined parameter.
            It is typical for the thickness of lobes to increase via an arithmetic progression, in each flow, such that the last lobe has the greatest thickness.
            \item This "initial" lobe has to be rasterized and added to the topography.
            See Section~\ref{step:raster} for details.
        \end{itemize}
        
        \subsection{Lobe propagation}
        In contrast to the initial lobes, new lobes are attached to previously deposited lobes. The creation of a new budding lobe can roughly be divided into two major tasks -- deciding the budding point, which is the point of attachment to the parent lobe, and calculating the dimensions of the new budding lobe.
        
        \subsubsection{Initial budding point}
        The procedure to select the initial budding point ensures that, on average, chains of lobes tend to follow paths of steepest descent. This procedure is described below. 
        \begin{itemize}
            \item First, the parent lobe is stochastically selected from all predecessor lobes; see Section S1 and S2 of the Supplemental Material for details about the rules for this selection.
            User-defined parameters can influence the tendency of lobes to bud close to or far from the vents.
            We denote the lobe center of the parent as $\vec{x}_{c, \text{parent}}$. 
            \item Once a parent lobe has been selected, a budding point, $\vec{x}_{b}$, on the boundary of the parent lobe is determined.
            This initial budding point, $\vec{x}_{b}$, is the point of minimum elevation along the circumference of the parent lobe. In the highly unlikely case that multiple such points exist, the tie-breaker rule outlined in \ref{appendix:tiebreaker} is used.
            \item The slope at $\vec{x}_{b}$ is calculated from the difference in elevation between the parent lobe center, $\vec{x}_{c, \text{parent}}$, and the budding point $\vec{x}_{b}$.
            \item The azimuthal angle of the child lobe, $\phi$, is set such that the semi-major axis points in the direction of $\vec{x}_{c, \text{parent}}-\vec{x}_{b}$.
        \end{itemize}
        
        \subsubsection{Adjusted budding point}
        The initial budding point is adjusted to account for uncertainty in the lobe propagation direction. This has the effect of introducing small random deviations from the path of steepest descent, which are correlated to the slope of the terrain. 
        This adjustment is summarized in the following. 
        
        \begin{itemize}
            \item The initial budding point, $\vec{x}_{b}$, can optionally be perturbed to obtain an adjusted budding point, referred to as angle perturbation in Figure~\ref{fig:cells_rasterized}(a).
            This step is the same as for initial lobes (described in Section~\ref{step:angle_perturbation}). Note that, as previously mentioned, this angle perturbation linearly increases for flatter terrains. 
            \item The effect of inertia can also be added to the budding lobe, to obtain an adjusted budding point, $\vec{x}_{b}'$.
            This inertial contribution can be understood as a reluctance of the flow to change direction.
            The new direction of the budding lobe semi-major axis is obtained as a weighted average of the semi-major axis directions of both the parent lobe and the budding lobe (see Section S2). 
        \end{itemize}
        
        \subsubsection{Budding lobe dimensions}
        As touched upon earlier, the slope of the terrain affects the shape of lobes. After the location of the budding point, $\vec{x}_{b}'$, has been determined, the dimensions of this lobe are calculated using the following:
        
        \begin{itemize}
            \item The slope is calculated from the elevation difference between $\vec{x}_{c, \text{parent}}$ and $\vec{x}_{b}'$.
            This is used to determine the semi axes, $a$ and $b$, of the new budding lobe, in the same way as is done for the "initial" lobes.
            \item The center of the budding lobe, $\vec{x}_c$, is calculated using the semi-major axis, $a$, and $\vec{x}_{c, \text{parent}}-\vec{x}_{b}'$.
            The default behaviour is to place the center such that the of the budding lobe has exactly one point of contact with its parent.
            \item The thickness, $t$, of the budding lobe is determined, previously elaborated in Section~\ref{step:thickness} (see also Section S2).
        \end{itemize}
        
        \subsection{Rasterization of Lobes} \label{step:raster}
        
        All lobes, regardless of their origin, have to be added to the topography.
        To do so, the lobes must be mapped onto the regular grid of the elevation data.
        This boils down to increasing the height of cells covered by the lobe, such that the added height is proportional to the covered fraction of the cell; i.e., rasterizing the lobe.
        Notably, this is one of the most expensive operations in the code.
        A key contribution is our novel, and efficient new strategy to rasterize lobes accurately (details in Section~\ref{sec:raster}).
         
        \subsection{Summary and Remarks}
        To recapitulate, every simulation is built up of flows, each of which is a chain of lobes. These lobes are progressively deposited on the topography. 
        
        Note that, since in each flow the thickness of constituent lobes either increases or decreases, the effect of multiple flows differs from a single flow of an equivalent total number of lobes. 
        For example, consider the case of one flow with 50 lobes, compared to 10 flows consisting of 5 lobes. In the former, there is only one lobe with the largest thickness. In the latter, since there are 10 chains of lobes, there are 10 lobes exhibiting the largest thickness.
        Furthermore, parent lobes are only selected within the same flow. Practically, this means that the simulation consisting of a single large flow, will, on average, spread out further from the vent.
        
        \emph{On a side note:} The inter-dependence of lobes and flows inhibits parallelization. Neither flows, nor lobes in a particular flow, can be processed in parallel due to race conditions resulting from this inter-dependence. On the other hand, simulations are independent of each other, and can therefore be amenable to "embarrassingly parallel" techniques of parallelization. Indeed, since the MrLavaLoba method~\cite{demichielivitturiMrLavaLobaNewProbabilistic2018} is stochastic in nature, an ensemble of runs should be aggregated (possibly in parallel) for statistical analysis.
        
        Currently, the lobe rasterization algorithm--arguably one of the most computationally intensive components of the method--does not leverage vectorization. This limitation arises because the most expensive part of the rasterization involves determining intersected cells, a process characterized by numerous branches and edge cases. Exploring vectorization for this algorithm is an intriguing direction, but one that we consider more suitable for future work.
        
        Finer points of the MrLavaLoba method of Vitturi \textit{et al.}~\cite{demichielivitturiMrLavaLobaNewProbabilistic2018}, as well as some important parameters are discussed in Section S1 of the Supplemental Material. For the sake of completeness, various formulas used in the method are also included in Section S2 of the Supplemental Material.
        
        \section{Algorithms}\label{sec:method} 
        
        In the following, key algorithms for our implementation of the MrLavaLoba \emph{method} of Vitturi \textit{et al.}~\cite{demichielivitturiMrLavaLobaNewProbabilistic2018} in Flowy are discussed.
        While this section provides a high-level overview of particularly important parts of the code, a large number of small optimizations have also been incorporated, all of which make their effect felt in the benchmarks and results shown later.
        
        \subsection{Lobe Rasterization}
        \label{sec:raster}
        The aim of rasterization is to convert each lobe (defined in Section~\ref{sec:model}) into thickness values on the square grid.
        For each cell, that is enclosed or intersected by the lobe, the relative fraction of the area lying in the interior of the ellipse is determined.
        The elevation value of that cell is then incremented by the area fraction times the thickness of the lobe. This means that intersected cells, which are cells that are not completely covered by the lobe, have smaller thicknesses than those of fully enclosed cells.
        It is noteworthy that this rasterization process is crucial for correctly modifying the input topography, thereby modelling the settling process of the lava.
        As touched upon earlier, lobe rasterization is one of the most expensive parts of the method.
        
        On a high-level, the lobe rasterization in Flowy proceeds as follows:
        \begin{enumerate}
            \item \textit{Bounding Box}:
            The bounding box is a rectangle, circumscribing the ellipse, with sides parallel to the rows and columns of the DEM grid (see~\ref{sec:bounding_box}).
            Since all cells, enclosed or intersected by the ellipse, are also enclosed by the bounding box, this helps to narrow down the candidates for rasterization. Figure~\ref{fig:cells_rasterized}(e) shows the extents of the bounding box of an ellipse as solid and dashed grey lines.
            As depicted in Figure~\ref{fig:cells_rasterized}(e), an intersected cell is a cell, which is passed through by the perimeter of the ellipse, while an enclosed cell is fully contained within the ellipse.
            Making this distinction is important, since it allows us to only perform the quadrature for the intersected cells (see Step~\ref{step:quad}).
            \item \textit{Cell Intersection}: 
            The next step is to determine if a cell is either fully enclosed by the ellipse, intersected by it, or outside of it, for all cells within the bounding box (refer to~\ref{sec:determining_enclosed_and_intersected_cells} for details). In order to achieve this, the bounding box is first spanned with horizontal line segments. 
            The intersected and enclosed cells can be determined from the intersection points of these line segments with the ellipse boundary.  
            \item \label{step:quad} \textit{Quadrature}: For the fully enclosed cells, no further work needs to be done -- since the area fraction is trivially equal to one.
            For the intersected cells, a numerical quadrature to determine the covered area fraction (which is less than one) is performed.
            This quadrature uses the trapezoidal rule.
            More details are provided in~\ref{sec:quad}.
            \item \textit{Volume Correction}): A certain numerical error is incurred every time a lobe is rasterized, leading to a numerical drift in the total deposited volume. To correct this drift, the volume actually deposited on the DEM is compared to the analytically known volume of the lobe.
            The difference is then distributed to the intersected cells (see~\ref{sec:volume_correction} for more information).
        \end{enumerate}
        
        \subsection{Hazard Map Generation}
        \label{sec:hazard_map}
        
        In essence, a hazard map indicates the likelihood of a region to be adversely affected by a volcanic eruption~\cite{loughlin2015global}.
        Intuitively, the hazard should be the greatest near the eruption site, and should shrink at the fringes of the lava flow. 
        
        In the MrLavaLoba method~\cite{demichielivitturiMrLavaLobaNewProbabilistic2018}, a qualitative measure of the hazard is calculated per flow, and added to cells of the topography.
        
        Similarly to the thickness, each lobe $k$ is assigned a qualitative hazard value $H(k)$, which is defined as
        \begin{equation}
            H(k) = C(k) + 1,
            \label{eq:hazard_lobe}
        \end{equation}
        where $C(k)$ is the number of cumulative descendants of the lobe $k$, given by the recursive definition
        \begin{equation}
            C(k) = \sum_{m \;\in\;\text{Ch}(k)} \left[C(m) + 1\right],
            \label{eq:cumulative_descendents}
        \end{equation}
        where
        \begin{equation}
            C(k) = 0 \text{   if   } \text{Ch}(k) = \emptyset,
        \end{equation}
        and $\text{Ch}(k)$ is a function that returns the set of children of the lobe $k$.
        
        The inspiration for this definition is the understanding that each time a lobe is added, a parcel of lava travels from the vent over all ancestor lobes.
        Thus, in this intuitive picture, the number of cumulative descendants corresponds to the number of times a parcel of lava has passed over a cell \cite{demichielivitturiMrLavaLobaNewProbabilistic2018}.
        
        To obtain a spatially resolved picture of the hazard, the hazard value of each lobe $k$, $H(k)$, must also be rasterized and associated to cells of the topography. This results in the so-called hazard map, denoted as $\mathcal{H}_{ij}$, where $(i,j)$ are the indices of a cell of the DEM.
        It is possible for a lobe and its parent to (partially) cover the same cell, therefore, to prevent over-counting, only the hazard due to the parent is associated with that cell.
        Consequently, the hazard value, $\mathcal{H}_{ij}$, is given by 
        \begin{equation}
            \mathcal{H}_{ij} = \sum_{\text{flows}} \sum_{k \in \text{lobes}} \sum_{\substack{(i,j) \in \\ \text{Cells}(k)}} \mathcal{\eta}^k_{ij},
            \label{eq:hazard_grid}
        \end{equation}
        where the first sum runs over all the flows in the simulation, the second sum runs over all lobes in the current flow and the last sum runs over all the cells in the grid.
        
        The indicator function $\eta^k_{ij}$ is defined as 
        \begin{equation}
            \eta^k_{ij} = 
            \left\{\begin{matrix}
                H(k) & \text{if} & (i,j) \notin \text{Cells}\big(\text{Par}(k)\big)\\
                0 & \text{else} &
            \end{matrix}\right. ,
        \end{equation}
        where $\text{Par}(k)$ denotes a function returning the parent of the lobe $k$ and $\text{Cells}(k)$ is a function returning the set of enclosed, or intersected, cells of the lobe $k$. 
        
        Evaluating the hazard map can become a time-consuming operation, and we have optimized the hazard map generation to minimize its cost.
        In the following, we outline the steps involved
        \begin{enumerate}
            \item \emph{Cumulative descendents}: The hazard map generation requires the number of cumulative descendants, $C(k)$, for each lobe $k$.
            However, it is not efficient to update $C(k)$ after the addition of each lobe (by iterating up the parent list). 
            Instead, we use a depth-first-search algorithm and make use of the recursive definition of $C(k)$, Eq.~\eqref{eq:cumulative_descendents}, to search through the tree after the completion of each flow.
            \item \emph{Hazard calculation}: Naturally, a performant implementation takes care to not evaluate the full sum over the grid [the third sum in \eqref{eq:hazard_grid}] for every single lobe.
            This is achieved by using a hashmap and caching the enclosed and intersected cells.
        \end{enumerate}
        
        For a discussion on probabilistic hazard maps, which are different from the qualitative hazard maps described above, see ~\ref{sec:probabilistic_hazard}.
        
        \section{Code Overview}\label{sec:code_overview}
        \label{subsec:code_structure}
        Flowy~\cite{flowy_github} has been written in modern object-oriented \texttt{C++20}.
        It is cross-platform, and binary executables can be compiled for Linux, MacOS and Windows machines.
        
        As touched upon in Section~\ref{sec:sim_structure}, the inter-dependence of lobes and flows within one run of Flowy makes parallelization over lobes, and even over flows, impractical.
        On the other hand, parallelization of separate runs of Flowy can be trivially accomplished.
        In practice, multiple parallel runs need to be analysed, due to the statistical nature of the method and it is trivial to leverage parallelism for this purpose.
        
        Further, we have also included Python bindings, using \texttt{pybind11}~\cite{pybind11} in the separate PyFlowy repository~\cite{pyflowy_github}.
        These bindings expose functions of Flowy to \texttt{Python}, enabling easy orchestration of Flowy runs with a well-known programmatic interface.
        PyFlowy is distributed through PyPI, using \texttt{cibuildwheel} to generate binary wheels for several host platforms, so that users do not even require a \texttt{C++} compiler for using Flowy through our \texttt{Python} bindings. Additionally, the source distribution can be compiled on any platform as it is provided via PyPI as well.
        
        In the following sections, we discuss inputs and outputs supported by Flowy.
        
        \subsection{Inputs}
        Flowy requires the following input files:
        \begin{enumerate}
            \item \emph{Configuration File}: User-defined information about the total erupted volume, vent locations and various simulation options should be provided in a configuration file, in the TOML format (see also, Section S1).
            The TOML file format is a well-known and popular configuration file format, designed to be human-readable and easy to parse in several languages.
            \item \emph{Topography File}: 
            The topography, upon which lobes will be deposited during the course of the simulation, must be provided to Flowy.
            As described in Section~\ref{sec:model}, the topography should be in the form of a regular two-dimensional grid with square cells.
            Flowy supports Esri ASCII files and binary NetCDF files~\cite{rewNetCDFInterfaceScientific1990}.
        \end{enumerate}
        
        \subsection{Outputs}
        Large-scale and long-term hazard maps can be invaluable for forecasting lava inundation, as well as for land-use planning and management (see also Section~\ref{sec:hazard_map}).
        However, the creation of long-term hazard maps typically requires many simulations, at which point data storage is a potential bottleneck. Flowy aims to alleviate this issue by incorporating a variety of size-reduction strategies for supported output formats. In this section, we discuss these output formats, as well as implemented options to reduce their size without compromising accuracy.
        
        Flowy supports a few standard output formats: the Esri ASCII data format, as well as binary NetCDF \cite{rewNetCDFInterfaceScientific1990} files.
        The ASCII grid format implemented in Flowy saves the floating point numbers with up to 6 significant digits, while omitting trailing zeros.
        This number of digits, roughly, corresponds to the precision of 32 bit floating point numbers.
        
        On the other hand, NetCDF files can also be optionally compressed. Since Flowy works with double precision (64 bit) floating point numbers, there is potential for lossy compression by packing the output data into either 32 bits or 16 bits.
        
        Additionally, the compression algorithms, implemented in the NetCDF library can be applied to leverage lossless compression.
        Further, we have implemented an optional crop-to-content feature, shaving off zeros (which represent "empty" cells) in thickness and hazard output files.
        All of these strategies can significantly reduce the size of output files.
        We quantitatively demonstrate their efficacy on a typical file output generated in a large scenario (see Table~\ref{tab:size} and Figure~\ref{fig:memory}). Compared to the ASCII grid format with 6 significant digits, a size reduction of the output files by a factor of 10 can be achieved without compromising accuracy (32 bit, compressed NetCDF output). A reduction by a factor of 20 can be achieved by leveraging 16 bit NetCDF output files. Perhaps somewhat counter-intuitively, the uncompressed binary NetCDF files are often larger than the ASCII files. The reason for this is the abundance of zeros, which are represented with only a single character in the ASCII file, but are saved with full precision in the binary files.
        
        Notably, the crop-to-content feature is the most beneficial for the reduction of output file sizes, when lossless compression is not used. The reason for that is that the lossless compression is highly efficient for repeated zeros, which are exactly what the crop-to-content function removes. Nonetheless, for any kind of further post-processing the compressed files have to be de-compressed, at which point cropping away the empty cells is worthwhile in any case.
        
        \section{Results and Discussion}\label{sec:results} 
        
        \subsection{Performance Comparison}
        
        \begin{figure}[t]
            \centering
            \includegraphics[width=\linewidth]{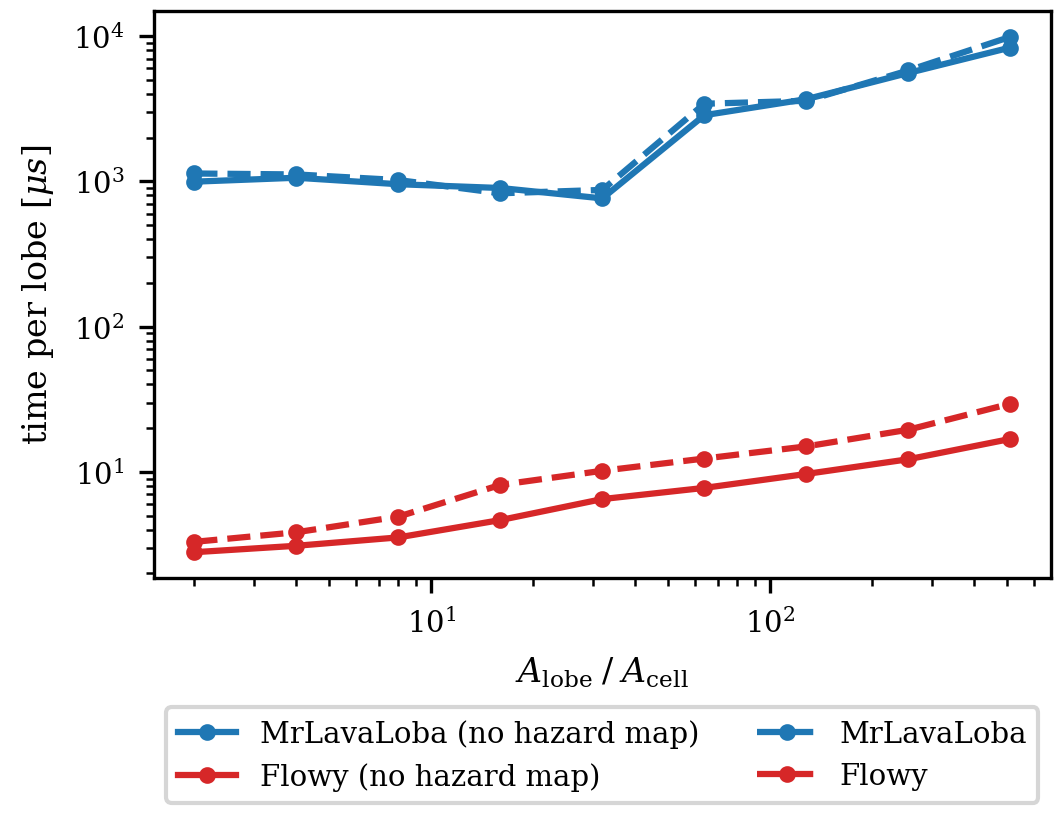}
            \caption{Comparison of the simulation run time, divided by the total number of deposited lobes, for Flowy (red) and the MrLavaLoba code (blue), with different lobe sizes.
            The topography used was a parabolic basin, which was gradually filled up with lobes.}
        \label{fig:performance_comparison}
        \end{figure}
        
        All comparisons of Flowy and the MrLavaLoba code were performed on an Intel(R) Xeon(R) Gold 6248R CPU @ 3.00 GHz, in single-threaded runs. 
        
        Figure~\ref{fig:performance_comparison} shows how Flowy scales with the relative increase in lobe size measured as the ratio of the area of the lobe ellipse divided by the area of the grid cells, $A_{\text{lobe}}/A_{\text{cell}}$. The specific shape of the topography does not directly influence the runtime. However, the number of lobes actually deposited can differ from the user-prescribed number, if flows reach the boundary of the topography. To alleviate this issue, a simple parabolic basin was used. In order to create DEMs with different cell sizes, the parabolic elevation function [of the form $h(x,y) = \alpha(x^2 + y^2)$] was sampled with different resolutions.
        
        The speedup of Flowy can be between 85 to 340 times, compared to the MrLavaLoba code (if the hazard map is created). In practice, simulations are usually run in the regime of $A_{\text{lobe}}/A_{\text{cell}}\approx 10$, where the speedup is about 200 to 300 times. 
        
        Clearly, generating hazard maps can be expensive, as is evident from Figure~\ref{fig:performance_comparison} (see also, Section~\ref{sec:hazard_map}).
        Runs of Flowy, in which the hazard map is \emph{omitted}, are between 1.2 to 1.7 times faster than those in which it is generated. This ratio increases with the ratio $A_{\text{lobe}}/A_{\text{cell}}$. 
        
        \subsection{Case Study: Eruption scenario in Geldingadalir valley}
        \label{sec:case_study}
        
        \begin{figure}[t]
            \centering
            \includegraphics[width=\linewidth]{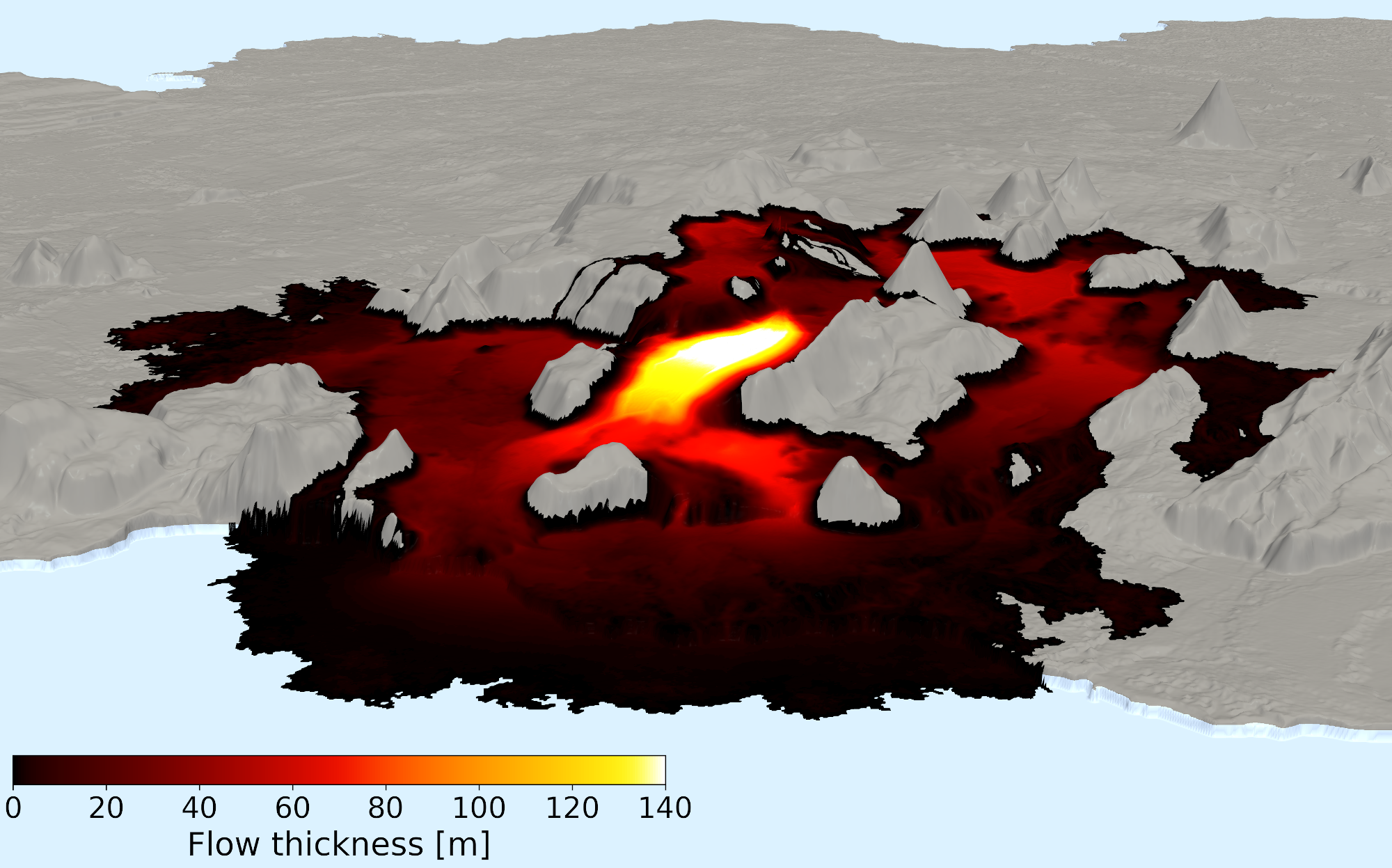}
            \caption{Three-dimensional view of a large-scale eruption scenario starting from the Geldingadalir valley.
            The lava flow thickness was calculated from $60$ runs of Flowy; see Section~\ref{sec:case_study} and Section S4 in the Supplemental Material for details. The ocean is depicted in light-blue, and the terrain is in grey.
            A perceptually uniform colormap was used for the lava thickness~\cite{kovesiGoodColourMaps2015}.}
        \label{fig:fagra_scenario}
        \end{figure}
        We performed $60$ runs of Flowy~\cite{flowy_github} and the MrLavaLoba code~\cite{lava_loba_github}, starting from a vent located in the Geldingadalir valley. Certain parameters were obtained from Pedersen \textit{et al.}~\cite{pedersenLavaFlowHazard2023}. However, this scenario is not meant to model the Fagradalsfjall eruption in 2021~\cite{pedersenLavaFlowHazard2023}, and serves only as a qualitative case study mimicking some features of this eruption.
        All used input parameters are listed in Table S1.
        The input DEM can be obtained from Zenodo~\cite{dem_2024}, and more information about it can be found in Section S2. 
        Figure~\ref{fig:fagra_scenario} shows a three-dimensional snapshot of the simulated scenario; depicting the averaged flow thickness calculated from the Flowy simulations.
        This is described in more detail below. 
        
        In order to obtain quantitative points of comparison, we computed the average maximum thickness $\overline{\max(h)}$, the average mean thickness $\overline{\langle h\rangle}$, the average covered area $\overline{A}$ and the standard deviations \eqref{eq:std_total} of those averages.
        For an exact definition of these quantities, see Equations~\eqref{eq:avg_max_thickness}, \eqref{eq:avg_thickness_mean} and \eqref{eq:avg_area} respectively.
        Lastly, to compare the outputs of the two codes in a spatially resolved manner, the average thicknesses $\overline{h_{ij}}$, per cell $(i,j)$ of the digital elevation model (DEM), and the standard deviations thereof $\sigma_{\overline{h_{ij}}}$ were determined. For definitions of these quantities, see Equations~\eqref{eq:avg_thickness_cell} and \eqref{eq:std_thickness_cell}.
        
        \begin{table*}[t]
        \centering
        \caption{Averaged quantities for 60 runs of Flowy and the MrLavaLoba code for the Geldingadalir valley case study.
        The average values and the standard deviations of the average are reported in the format $\overline{x} \pm \sigma_{\overline{x}}$.
        The averages were computed for the full thickness, as well as the volume masked thickness (threshold=0.96).}
        \label{tab:averaged_quantities}
        \begin{tabular}{l c r r r r r r}
        \toprule
            & & \multicolumn{2}{c}{Full} & \multicolumn{2}{c}{Masked} \\
          Quantity & Unit & \multicolumn{1}{c}{Flowy} & \multicolumn{1}{c}{MrLavaLoba} & \multicolumn{1}{c}{Flowy} & \multicolumn{1}{c}{MrLavaLoba}\\
        \midrule
        $\overline{\max(h)}$ & [m]           & 160.5 $\pm$ 1.2  & 161.4 $\pm$ 1.10  & 160.5 $\pm$ 1.2 & 161.4 $\pm$ 1.1 \\
        $\overline{\langle h \rangle}$ & [m] & 21.2 $\pm$ 0.12 & 21.1 $\pm$ 0.087 & 30.6 $\pm$ 0.16 & 30.7 $\pm$ 0.13\\
        $\overline{A}$ & [km$^2$]            & 0.246 $\pm$ 0.0014  & 0.246 $\pm$ 0.0010  & 0.1633 $\pm$ 0.0009  & 0.1629 $\pm$ 0.0007 \\
        \bottomrule
        \end{tabular}
        \end{table*}
        
        \begin{figure}[t]
            \centering
            \includegraphics[width=\linewidth]{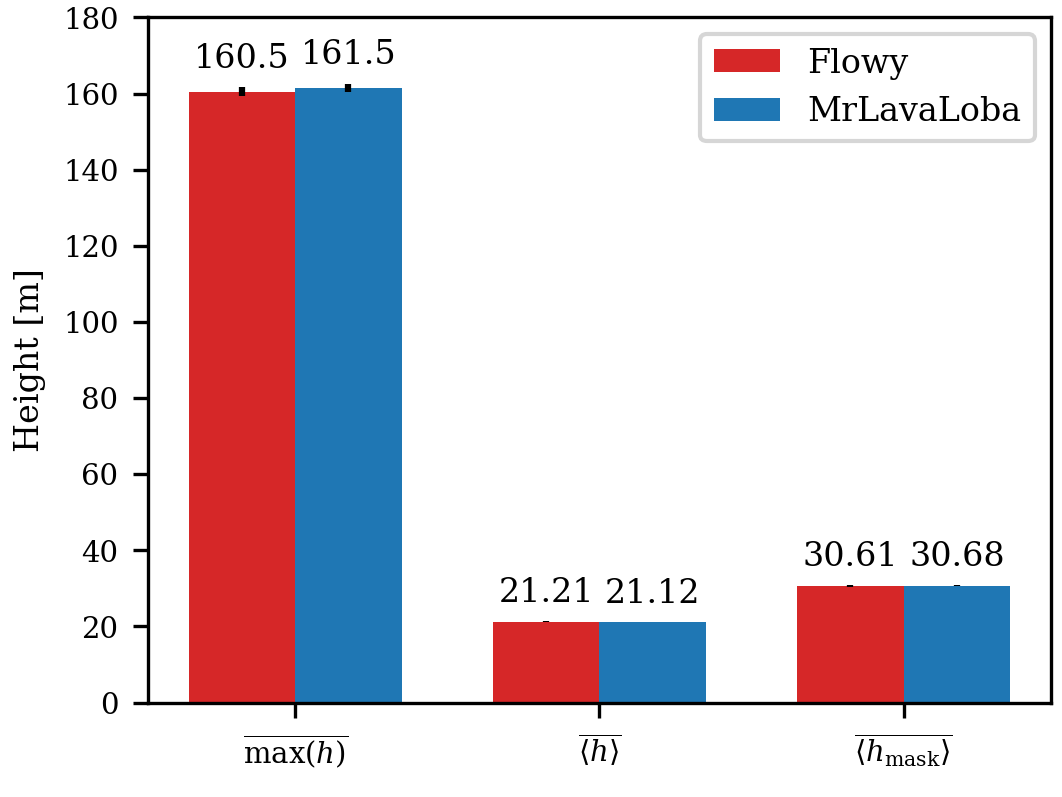}
            \caption{Comparison of averaged thickness quantities for the Geldingadalir valley case study (from left to right): Maximum thickness, mean thickness, mean masked thickness (threshold=0.96).
            Red bars are for the Flowy code; blue bars are for the MrLavaLoba code.
            The numbers on top of the bars are the exact numerical values.}
        \label{fig:averaged_quantities}
        \end{figure}
        
        \begin{figure}[t]
            \centering
            \includegraphics[width=\linewidth]{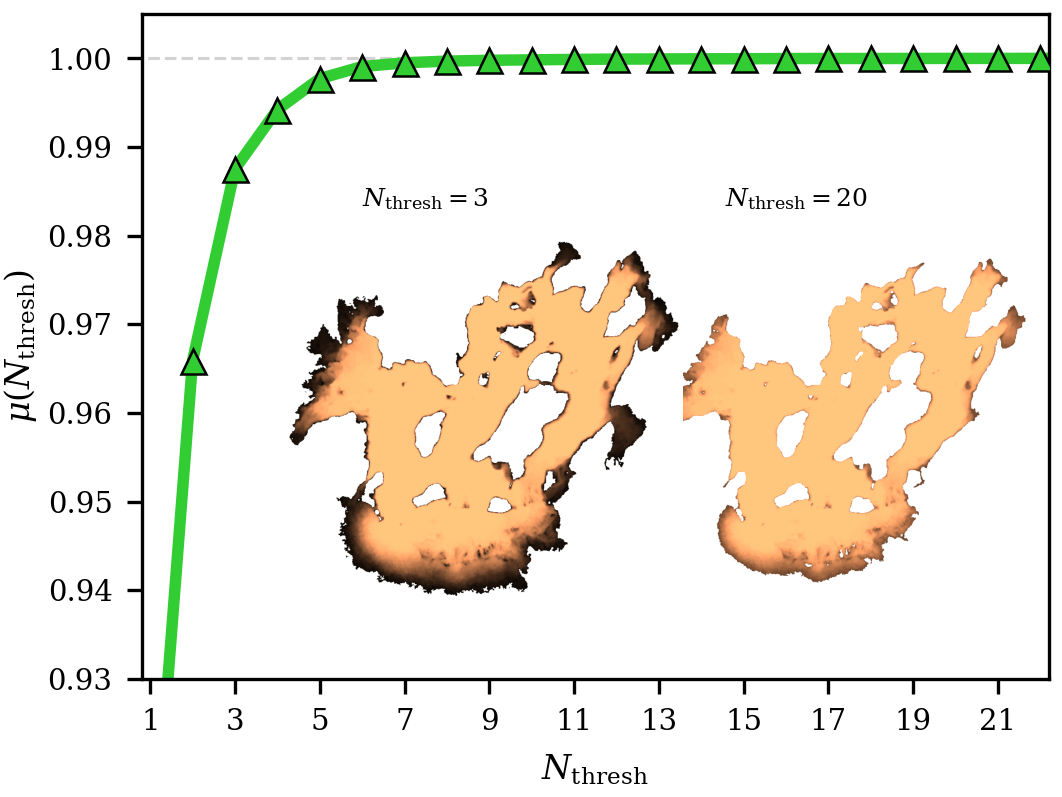}
            \caption{Trend of the filtered coverage convergence index, $\mu (N_{ \text{thresh}})$, with the sampling count threshold, $N_{\text{thresh}}$.
            A $N_{\text{thresh}}$ value of 3 means that all cells which are touched by less than 3 runs are discarded.
            A dashed line denotes $\mu (N_{ \text{thresh}})=1$, representing perfect agreement. 
            Insets show the sampling counts obtained in the Flowy simulations, when filtered by $N_{\text{thresh}} = 3$ and $N_{\text{thresh}} = 20$. The same color scheme is used for the insets as in Figure~\ref{fig:comparison_counts}.
            }
        \label{fig:threshold_count_convergence}
        \end{figure}
        
        The results of the comparisons between the average quantities are tabulated in Table~\ref{tab:averaged_quantities} and (partially) visually represented in Figure~\ref{fig:averaged_quantities}.
        Agreement, within statistical margins, was reached for all. On average one run of Flowy took $\approx 10$ seconds, whereas one run of the MrLavaLoba code took $\approx 1380$ seconds (23 minutes).
        Therefore, Flowy is around \emph{140 times faster} on this example.
        In addition, we also quantified the effects of various size reduction options in the output files for this case study, listed in Table~\ref{tab:size}.
        
        Renderings of the averaged thicknesses $\overline{h_{ij}}$ are to be found in Fig.~\ref{fig:comparison_thickness_full} and for the masked thickness (defined in~\ref{appendix:masked_thickness}) in Fig.~\ref{fig:comparison_thickness_masked}; both show excellent agreement, with minor differences towards the edges of the flows, where individual cells are under-sampled.
        The sampling of cells in independent simulation runs can be quantified by the sampling count, $N_{ij}^{+}$, which is defined as the number of times each cell was touched by lava in $N$ simulations [see ~\eqref{eq:n_nonzero_runs}].
        Naturally, the average thickness tends to differ slightly in the under-sampled regions, as evidenced by Fig.~\ref{fig:comparison_counts}.
        
        The agreement between the average thickness and average masked thickness can be further quantified by a relative root mean-square (RMS) error.
        The RMS error has been extensively used to quantify differences between gray-scale images, and it is popular for the development of
        image compression algorithms~\cite{delpImageCompressionUsing1979,habibiHybridCodingPictorial1974,jainImageDataCompression1981,lindeAlgorithmVectorQuantizer1980,wilsonNewMetricGreyScale1997}.
        
        \begin{figure*}[t]
            \centering
            \includegraphics[width=\linewidth]{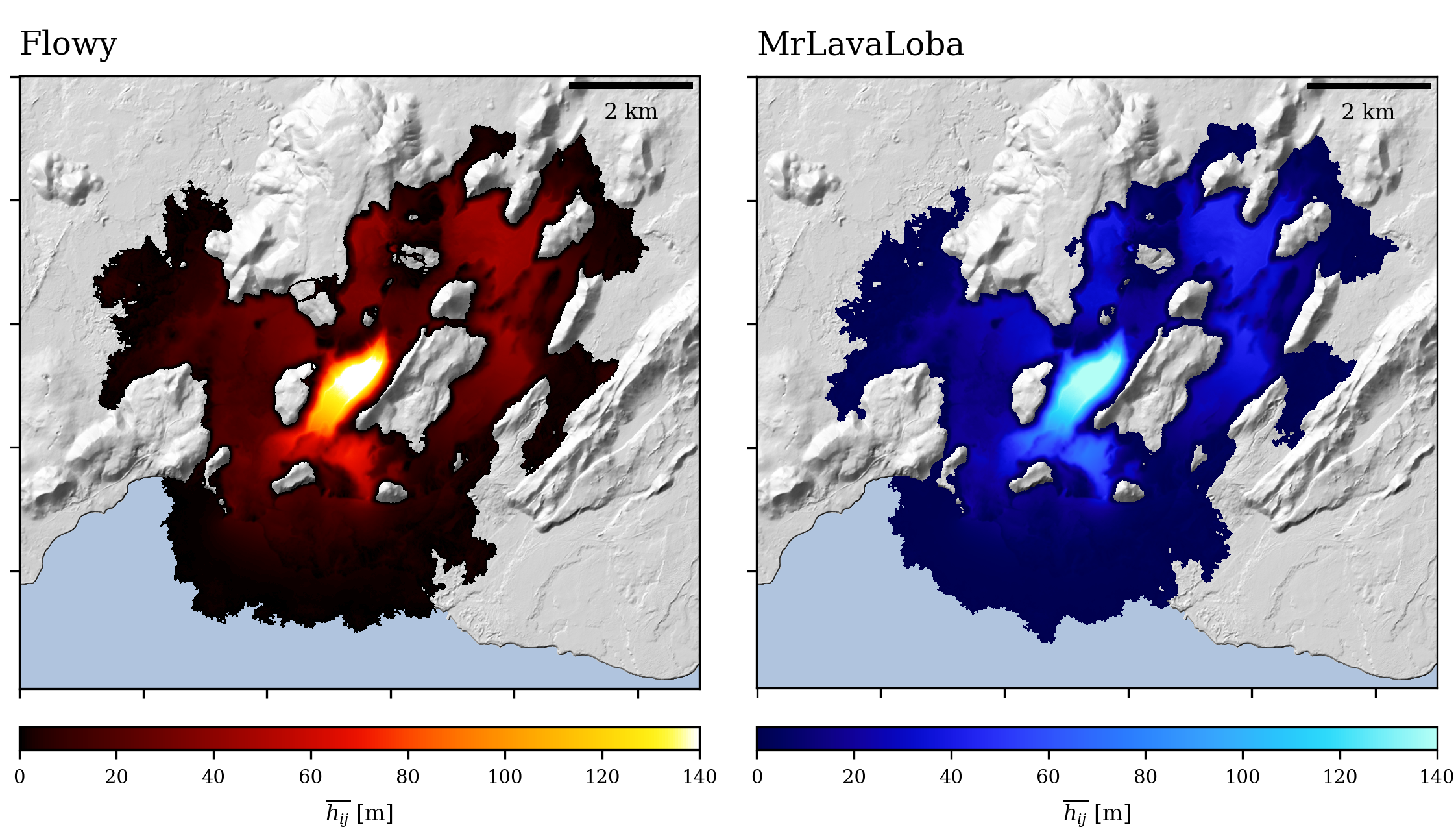}
            \caption{Comparison of the average cell resolved full thickness $\overline{h_{ij}}$, for the Flowy code (left, red) and the MrLavaLoba code (right, blue).
            The shade indicates the thickness of the lava deposit, with darker shades meaning thinner deposits.
            The ocean is light-blue, and the terrain is grey.}
            \label{fig:comparison_thickness_full}
        \end{figure*}
        
        \begin{figure*}[t]
            \centering
            \includegraphics[width=\linewidth]{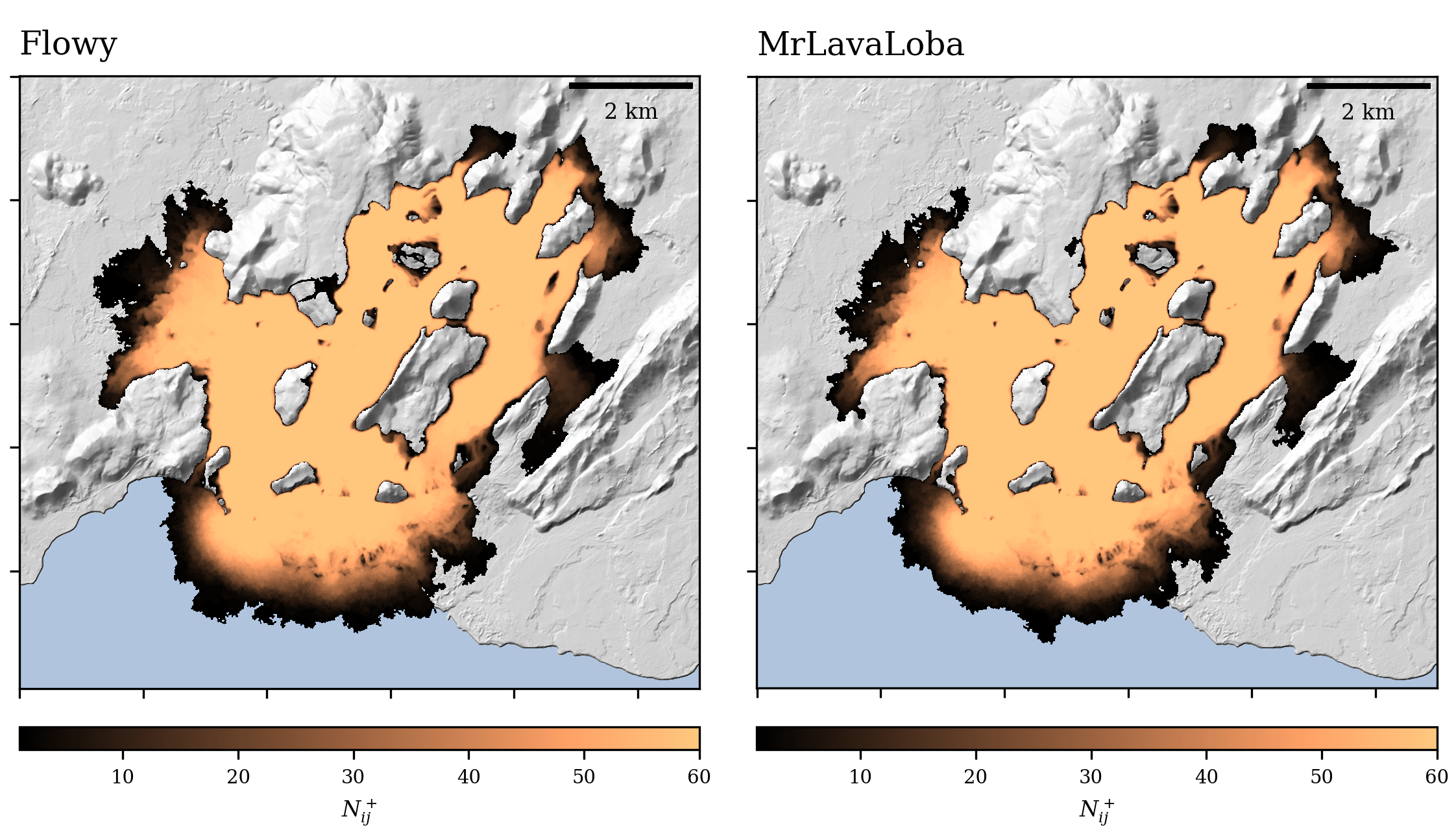}
            \caption{Comparison of the number of times, each cell was touched by lava summed over 60 runs ($N^+_{ij}$) of the Flowy code (left) and the MrLavaLoba Code (right), respectively.
            Copper colored regions correspond to a high number of flows touching a cell, while dark regions correspond to low numbers.
            Intuitively, this is connected to the local uncertainty in the inundated area.}
            \label{fig:comparison_counts}
        \end{figure*}
        
        The values of the relative RMS error, $\Delta_{\text{rel}}$, [defined in~\eqref{eq:rel_rms}] for the average thickness and masked thickness are $0.059$ and $0.046$, respectively.
        The values are consistent with the excellent agreement suggested by visual inspection of Figure~\ref{fig:comparison_thickness_full} and Figure~\ref{fig:comparison_thickness_masked} (also supported by the comparison of averaged quantities in Table~\ref{tab:averaged_quantities}). 
        
        To qualify the agreement between the two simulation outputs further, we compute the coverage convergence index $\mu$ [defined in \eqref{eq:coverage_index}].
        The coverage convergence was originally used to quantify the agreement between single simulation runs~\cite{demichielivitturiMrLavaLobaNewProbabilistic2018}, but here we calculate $\mu$ for the averaged thickness outputs of Flowy and the MrLavaLoba code.
        
        Further, to improve the statistical quality of the comparison,
        we remove cells with a low sampling count $N_{ij}^{+}$ (blacker regions in Figure~\ref{fig:comparison_counts}). The filtered covered convergence index, $\mu(N_{\text{thresh}})$, so obtained, is defined in~\eqref{eq:filtered_convergence_index}. 
        We show that this filtering can facilitate convergence of $\mu(N_{\text{thresh}})$ to $1$, as depicted in Figure~\ref{fig:threshold_count_convergence}. The convergence is rapid and a value of $\mu(N_{\text{thresh}}) > 0.99$ is already achieved for a threshold of $N_{\text{thresh}} = 4$ (the maximum possible value would be 60).

        

        
        
        \begin{table}[t]
        \caption{Summary of the effects of various size reduction strategies on the size of supported file formats.
        The files used are a typical output of a large scenario.
        Files can be cropped to content (thereby removing extraneous "empty cell" values in thickness and hazard output maps), and can be compressed via lossless compression and lossy compression.
        The maximum and minimum sizes in this example are highlighted. Note that the binary uncompressed NetCDF files can be quite large, and can often be larger than the ASCII files.}
        \label{tab:size}
        \begin{tabular}{@{}lccr@{}}
        \toprule
        File type      & \begin{tabular}[c]{@{}c@{}}Compression\\ level\end{tabular} & \begin{tabular}[c]{@{}c@{}}Crop to\\ content\end{tabular} & Size (MB)       \\
        \midrule[0.3pt]
        ASCII grid     & -                                                           & No              & 32.02           \\
        ASCII grid     & -                                                           & Yes             & 11.56           \\
        \midrule[0.1pt]
        NetCDF, 64 bit & 0                                                           & No              & \textbf{110.49} \\
        NetCDF, 64 bit & 0                                                           & Yes             & 28.62           \\
        NetCDF, 64 bit & 9                                                           & No              & 4.99            \\
        NetCDF, 64 bit & 9                                                           & Yes             & 4.81            \\
        NetCDF, 32 bit & 0                                                           & No              & 55.28           \\
        NetCDF, 32 bit & 0                                                           & Yes             & 14.33           \\
        NetCDF, 32 bit & 9                                                           & No              & 2.45            \\
        NetCDF, 32 bit & 9                                                           & Yes             & 2.34            \\
        NetCDF, 16 bit & 0                                                           & No              & 27.67           \\
        NetCDF, 16 bit & 0                                                           & Yes             & 7.18            \\
        NetCDF, 16 bit & 9                                                           & No              & 1.27            \\
        NetCDF, 16 bit & 9                                                           & Yes             & \textbf{1.19}  \\
        \bottomrule
        \end{tabular}
        \end{table}
        
        \begin{figure}[t]
            \centering
            \includegraphics[width=\linewidth]{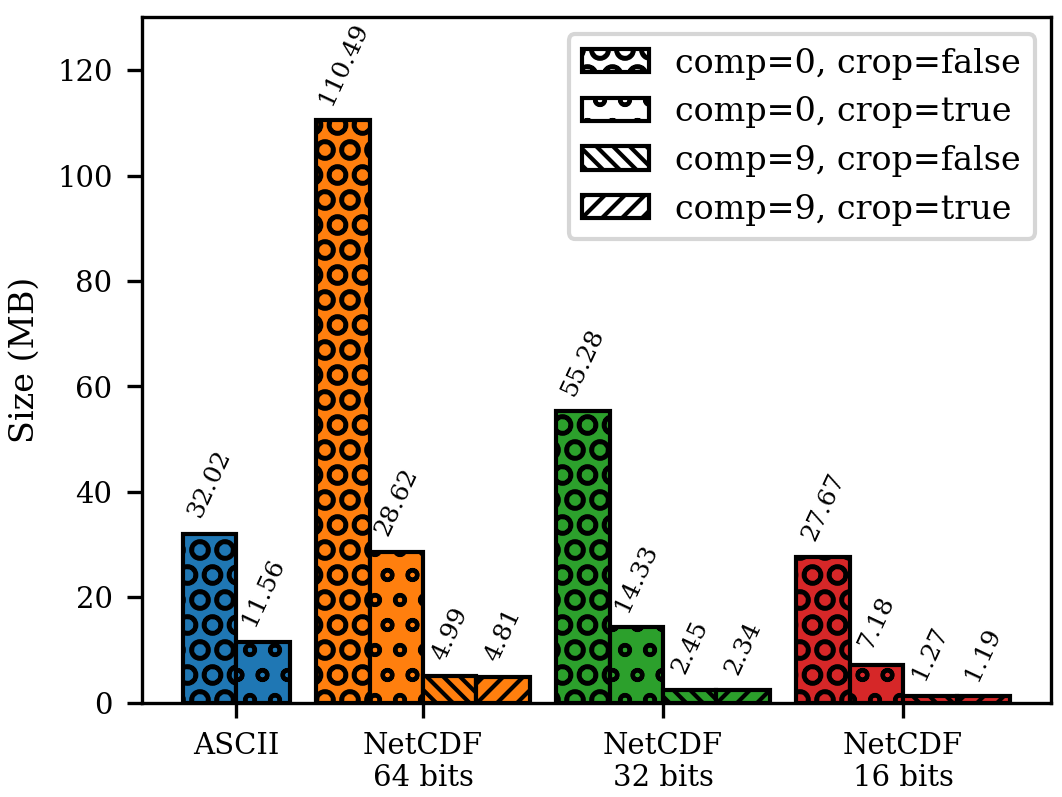}
            \caption{Comparison of the efficacy of various size reduction strategies on the size of supported file formats. 
            The legend shows the compression level, or comp, and whether the files are cropped to content or not.}
        \label{fig:memory}
        \end{figure}
        
        \section{Conclusions}\label{sec:conclusions}
        Flowy is a new code, that implements the MrLavaLoba method -- a proven probabilistic method for lava emplacement prediction, pioneered by Vitturi \textit{et. al.} ~\cite{demichielivitturiMrLavaLobaNewProbabilistic2018}.
        Flowy targets two major bottlenecks in large-scale lava simulations--runtime and data storage. We have designed and implemented fast algorithms for critical parts of the method. In particular, we have implemented a fast, efficient and accurate algorithm to rasterize lobes onto the regular grid of the topography, which is an expensive operation. 
        Further, we have proposed and implemented a scheme for volume correction, which ensures that there is no drift from the user-prescribed eruption volume.
        
        Another time-consuming operation is the hazard map generation. To optimize this, we have implemented an efficient scheme for generating the hazard map, based on depth-first recursion. 
        
        The performance of Flowy has been demonstrated for a parabolic basin, which generalizes to arbitrary topographies. For typical use cases, the speedup, when compared to the the established and well-used MrLavaLoba reference code, can be expected to lie between 130 to 200 times.
        
        Achievable reductions in output file size have been demonstated (up-to 10-20 times smaller), leveraging compression strategies and a new crop-to-content feature.  
        
        The implementation of the method has been verified in a realistic case study, by comparison to the MrLavaLoba code. All obtained outputs agree within statistical margins. Additionally, methods to obtain spatially resolved, statistically significant comparisons have been discussed and applied.
        
        We have provided Python bindings to Flowy, in PyFlowy, to facilitate accessibility. In the future, PyFlowy could be instrumental in creating a GUI (Graphical User Interface), making the workflow even more user-friendly. We also envisage the possibility of integration with the widely used QGIS~\cite{QGIS_software} software, which is a popular geographic information system and visualization tool. 
        
        Further, the modular design of Flowy allows for easy extensions, modifications and implementation of other methods of lava emplacement. 
        The algorithms presented here, could be applied to lava simulation methods with foundations in Computational Fluid Dynamics. Moreover, hybrid methods combining strengths of both approaches are an attractive avenue of future research.
        
        \section*{CRediT authorship contribution statement}
        \noindent\textbf{Moritz Sallermann \& Amrita Goswami}: Conceptualization, Methodology, Software, Validation, Formal analysis, Writing - Original Draft, Writing - Review \& Editing, Visualization\\
        \textbf{Alejandro Peña-Torres}: Conceptualization, Validation, Formal analysis, Writing - Original Draft, Writing - Review \& Editing\\
        \textbf{Rohit Goswami:} Software, Writing - Original Draft, Writing - Review \& Editing

        \section*{Declaration of competing interest}
        The authors declare that they have no known competing financial interests or personal relationships that could have appeared to influence the work reported in this paper.
        
        \section*{Acknowledgements}\label{sec:acknowledgements}
        This work was partially funded by the Icelandic Research Fund (grant 228615-051).
        The calculations were performed using compute resources provided by the Icelandic Research Electronic Infrastructure (IREI). 
        We are grateful to Hannes J\'{o}nsson for fruitful discussions, advice and support. The MrLavaLoba code was used as a reference while implementing the MrLavaLoba method in Flowy. 
        
        \appendix
        
        \section{Intersections of a line segment with an ellipse}
        \label{sec:line_segment_intersection}
        One elementary operation, used throughout the next steps of the rasterization, is the computation of the line segment intersections with an ellipse.
        The line segment $L$, between the endpoints $\vec{p}$ and $\vec{q}$, is the set of points given by 
        \begin{equation}
            L = \Big\{ \vec{p} + t \left(\vec{q} - \vec{p}\right)\;|\;t \in  [0,1] \Big\}.
            \label{eq:line_segment}
        \end{equation}
        
        We define the intersection points between the line segment $L$ and the ellipse \emph{as the first and the last point of the line segment that lie within the interior of the ellipse}.
        The point of this definition is that, if a line segment ends or starts within the interior of the ellipse, the respective start/endpoint(s) will be considered the point(s) of intersection.
        
        To compute the intersection points, according to the definition above, we first transform the problem into a coordinate system with axes aligned with the semi-axes of the ellipse and with the origin coinciding with its center.
        This is achieved by the affine transformation
        \begin{equation}
            \vec{x}' = 
            \begin{pmatrix}
                \cos \phi & \sin \phi\\
                -\sin \phi & \cos \phi
            \end{pmatrix} 
            \left( \vec{x} - \vec{x}_c \right).
        \end{equation}
        
        In this coordinate system, the equation for the ellipse perimeter $E$ is given by 
        \begin{equation}
           \vec{x} = (x_1, x_2) \in E \equiv \left(\frac{x_1}{a}\right)^2 + \left(\frac{x_2}{b}\right)^2 = 1
           \label{eq:perimeter}.
        \end{equation}
        
        Plugging \eqref{eq:line_segment} into \eqref{eq:perimeter}, leads to a quadratic equation for $t$ with the solutions
        \begin{equation}
            t_{1/2} = \frac{-\beta \mp \sqrt{\beta^2 - 4 \alpha \gamma}}{2\alpha},
        \end{equation}
        where 
        \begin{align*}
            \vec{d} &= \vec{q'} - \vec{p'},\\
            \alpha &= \begin{pmatrix}
                a^{-2}\\
                b^{-2}
            \end{pmatrix}
            \begin{pmatrix}
               d_1^2 & d_2^2
            \end{pmatrix},\\
            \beta &= 2 \begin{pmatrix}
                a^{-2}\\
                b^{-2}
            \end{pmatrix}
            \begin{pmatrix}
               {p'_1}d_1 & {p'_2}d_2
            \end{pmatrix},
        \end{align*}
        and
        \begin{align*}
            \gamma &= \begin{pmatrix}
                a^{-2}\\
                b^{-2}
            \end{pmatrix}
            \begin{pmatrix}
               {p'_1}^2 & {p'_2}^2
            \end{pmatrix} - 1.
        \end{align*}
        
        If the radicand, $\beta^2 - 4\alpha \gamma$, is smaller than zero, the line segment has no intersections with the ellipse and we are done.
        Otherwise, we have to check the overlap of the interval $[0,1]$ with the interval of solutions $[t_1, t_2]$.
        
        \emph{No intersection points exist} if 
        \begin{equation}
            [0,1] \cap [t_1, t_2] = \emptyset,
            \label{eq:no_intersection_points}
        \end{equation}
        or, equivalently,
        \begin{equation}
          (t_1 < 0 \land t_2 < 0) \lor (t_1 >1 \land t_2 > 1).
        \end{equation}
        If \eqref{eq:no_intersection_points} is not fulfilled, we compute the intersection points by plugging 
        $t_\text{first} = \min\left(0, t_1\right)$ and $t_\text{last} = \max\left(1, t_2\right)$ into \eqref{eq:line_segment}.
        
        \section{The square grid}
        \label{sec:grid}
        Many of the computations in Flowy concern quantities defined on a square lattice, with spacing $s$.
        Frequently, this will be an elevation or a flow thickness.
        Each cell of the lattice is defined by a pair of indices $(i,j)$, where the index $i$ is associated with the $x$-coordinate and the index $j$ is associated with the $y$-coordinate.
        We denote quantities, defined on the lattice, with subscripts for the cell indices, e.g. an elevation at the coordinates $x_i$ and $y_i$ could be written as $h_{ij}$.
        Generally, the coordinates $x_i$ and $y_i$ are to be understood as the lower left corner of the cell.
        This means a point defined by the cartesian coordinates $(x,y) \in \mathbb{R}^2$ can be associated to the cell indices $(i,j) \in \mathbb{Z}^2$ by
        \begin{equation}
            (i,j) = \left( \left\lfloor \frac{x}{s} \right\rfloor, \left\lfloor \frac{y}{s} \right\rfloor \right),
            \label{eq:coord_to_grid}
        \end{equation}
        where $s$ is the spacing of the square lattice.
        
        \section{Bounding Box of a Lobe}
        \label{sec:bounding_box} 
        By the bounding box of a lobe we refer to the smallest rectangular subsection of the square lattice that fully contains a circumscribing rectangle of the ellipse, with sides parallel to the cardinal directions of the lattice. 
        Fortunately, this concept is rather easy to appreciate visually [see Figure~\ref{fig:cells_rasterized} (e)].
        
        Let the circumscribing rectangle $R_\text{circ}$ be the set
        \begin{equation}
            R_\text{circ} = \Big\{ (x,y) \;\big|\; x \in [x_1, x_2] \land y \in [y_1, y_2]\Big\} \subset \mathbb{R}^2.
        \end{equation}
        Then, the closed intervals $[x_1, x_2]$ and $[y_1, y_2]$ can be determined from the center of the ellipse $\vec{x}_c = (x_c, y_c)$, the azimuthal angle $\phi$ and the semi-axes $a$ and $b$ by
        \begin{align}
            x_{1/2} &= x_c \mp \sqrt{a_x^2 + b_x^2},\\
            y_{1/2} &= y_c \mp \sqrt{a_y^2 + b_y^2},
        \end{align}
        where
        \begin{equation*}
            \begin{pmatrix}
                a_x & b_x\\
                a_y & b_y
            \end{pmatrix} = 
            \begin{pmatrix}
                \cos \phi \\
                \sin \phi
            \end{pmatrix}
            \begin{pmatrix}
                a & b
            \end{pmatrix}
            .
        \end{equation*}
        The endpoints of these two intervals in $x$ and $y$ can then be transformed into cell indices $(i_1, j_1)$ and $(i_2, j_2)$ by using Eq.~\eqref{eq:coord_to_grid}.
        Finally, the bounding box is the set 
        \begin{equation}
            \mathcal{B} = \Big\{ (i,j) \;\big|\; i \in [ i_1, i_2] \land j \in [j_1, j_2]\Big\} \subset \mathbb{Z}^2.
        \end{equation}
        
        \section{Determining enclosed and intersected cells}
        \label{sec:determining_enclosed_and_intersected_cells}
        Once the bounding box has been determined, we find the cells that are intersected, or enclosed, by the lobe by scanning the bounding box with line-segments, as depicted in Figure~\ref{fig:cells_rasterized}(d) and Figure~\ref{fig:cells_rasterized}(e).
        As stated in the main text: an intersected cell is a cell, which is passed through by the perimeter of the ellipse, while an enclosed cell is fully contained within the ellipse.
        Making this distinction is crucial, since it allows us to only perform the quadrature for the intersected cells.
        
        For each row of cells $j \in \left[ j_\text{min},j_\text{max} \right]$, within the bounding box, we compute the intersection of the lobe ellipse with two line segments: 
        \begin{enumerate}
            \item One segment along the bottom of the row, from $\vec{p} = (x_{i_1}, y_{j})$ to $\vec{q} = (x_{i_2}, y_{j})$
            \item and one along the top of the row, from $\vec{p} = (x_{i_1}, y_{j+1})$ to $\vec{q} = (x_{i_2}, y_{j+1})$.
        \end{enumerate}
        Thus, we get up to four intersection points: two points to the left and two points to the right, with one at the bottom of the row and another at the top.
        
        After transforming these intersection points into cell indices by using \eqref{eq:coord_to_grid}, we obtain four different $i$ indices: $i^\text{left}_\text{bot}$, $i^\text{left}_\text{top}$, $i^\text{right}_\text{bot}$ and $i^\text{right}_\text{top}$. 
        The $j$ index is, of course, constant since we are scanning along lines with a fixed $y$-coordinate.
        From these four indices, the intersected and enclosed cells within $j$-th row can be found.
        The intersected cells in row $j$, $\mathcal{I}_{j}$, are
        \begin{equation}
            \mathcal{I}_{j} = \Big\{ (i,j) \;\big|\; i \in \left[ i^\text{left}_\text{start}, i^\text{left}_\text{stop} \right] \cup \left[  i^\text{right}_\text{start}, i^\text{right}_\text{stop} \right]  \Big\},
        \end{equation}
        and the enclosed cells, $\mathcal{E}_{j}$, are
        \begin{equation}
            \mathcal{E}_{j} = \Big\{ (i,j) \;\big|\; i \in \left[ i^\text{left}_\text{stop} + 1, i^\text{left}_\text{start} - 1 \right]   \Big\},
        \end{equation}
        where 
        \begin{align*}
            &i^\text{left}_\text{start} = \min \left( i^\text{left}_\text{bot}, i^\text{left}_\text{top} \right)\\
            &i^\text{left}_\text{stop} =  \max \left( i^\text{left}_\text{bot}, i^\text{left}_\text{top} \right)\\
            &i^\text{right}_\text{start} = \min \left( i^\text{right}_\text{bot}, i^\text{right}_\text{top} \right)
        \end{align*}
        and
        \begin{align*}
            &i^\text{right}_\text{stop} =  \max \left( i^\text{right}_\text{bot}, i^\text{right}_\text{top} \right).
        \end{align*}
        
        Finally, the total set of intersected and enclosed cells is found by taking the unions
        \begin{equation}
            \mathcal{I} = \cup_{j
            } \mathcal{I}_{j} 
        \end{equation}
        and
        \begin{equation}
            \mathcal{E} = \cup_{j} \mathcal{E}_j.
        \end{equation}
        
        Care must be taken in a few edge cases:
        \begin{enumerate}
            \item If $i^\text{left}_\text{stop} = i^\text{right}_\text{start}$ (which is rare but can happen if the tip of an ellipse barely overlaps with a row) $i^\text{right}_\text{start}$ has to be incremented by 1 to avoid double counting.
            \item For the top/bottom row of cells of the bounding box, the top/bottom of the row will have no intersections with the ellipse and none of the cells will be enclosed.
            The reason for this is the definition of the bounding box. The intersected cells at the top/bottom row are 
            \begin{equation}
                \mathcal{I}_{j} = \Big\{ (i,j) \;|\; i \in \left[ i^\text{left}_\text{bot/top}, i^\text{right}_\text{bot/top} \right] \Big\}.
            \end{equation}
        \end{enumerate}
        
        \section{Quadrature within Cells}
        \label{sec:quad}
        For the intersected cells, determined in \ref{sec:determining_enclosed_and_intersected_cells}, a numerical quadrature is performed to assess the fraction of the cell area covered by the ellipse. For the enclosed cells, however, this step can be skipped, since the fraction equals 1 by definition.
        
        Let a particular cell, $C$, be defined by the set of points 
        \begin{equation}
            C = \Big\{ (x,y) \;\big|\; x \in [x_\text{l}, x_\text{r}] \land y \in [y_\text{b}, y_\text{t}]\Big\} \subset \mathbb{R}^2.
        \end{equation}
        To assess the fractional area of $C$, which is covered by the lobe ellipse, the cell is divided into a finite number of rows with fixed $y$ coordinates
        \begin{equation}
            y_i = y_\text{b} + \frac{i-1}{N-1} (y_\text{t} - y_\text{b}) \text{  for  } i \in [1,N],
        \end{equation}
        where $N$ is the number of rows.
        
        For each row, the two intersections points between the line segment, with endpoints $\vec{p_i} = \left(x_\text{l}, y_i\right)$ and $\vec{q_i} = \left(x_\text{l}, y_i\right)$, and the ellipse are computed, according to \ref{sec:line_segment_intersection}.
        
        \textit{Please note:} This is a slight simplification to aid readability. In practice, these intersections are computed at the same time as the cells are determined (\ref{sec:determining_enclosed_and_intersected_cells}), in order to not perform redundant calculations of line segment intersections.
        
        For each row $i$, let the intersection point on the left of the cell be denoted by 
        \begin{equation}
            \vec{u}_1 = \left(l_i, y_i\right)
        \end{equation}
        and the one on the right of the cell by 
        \begin{equation}
            \vec{u}_2 = \left( r_i, y_i \right).
        \end{equation}
        Then, the fraction of the covered cell area, $f_\text{cell}$, can be approximated with the trapezoidal rule as
        \begin{equation}\label{eq:frac}
            f_\text{cell} \approx \frac{1}{2 (N-1) s} \sum_{i=0}^{N-1} r_{i+1} - l_{i+1} + r_{i} - l_{i},
        \end{equation}
        where $s$ is the spacing of the square grid, as defined in \ref{sec:grid}.
        We note that, since ellipses are convex shapes, the trapezoidal rule tends to systematically underestimate $f_\text{cell}$. This is addressed in the following section.
        
        \section{Volume correction}
        \label{sec:volume_correction}
        
        During the rasterization process, the covered area fraction of the intersected cells is calculated using a trapezoidal quadrature rule, as described in Section~\ref{sec:quad}. 
        However, the nature of this integration necessarily implies that the area is underestimated.
        As more and more lobes are rasterized, these errors do not cancel out, and there is numerical drift from the prescribed erupted volume.
        We note that the MrLavaLoba code also exhibits numerical drift from the erupted volume, albeit in the opposite direction.
        
        We have implemented an optional volume correction, in Flowy. 
        This relies on the fact that the true volume of the lobe is known exactly.
        Therefore, the deficit in volume, which must be added so that the true volume is deposited on the topography, is defined as
        \begin{equation}
            \Delta V = \pi a b t - V_{\text{raster}},
        \end{equation}
        where $a$ and $b$ are the lobe semi-axes, $t$ is the thickness of the lobe, and $V_{\text{raster}}$ is the volume obtained by rasterizing the lobe. 
        Subsequently, this extra volume is added to the intersecting cells, weighted by the covered fraction of each cell. 
        Therefore, for a particular intersecting cell $(i, j)$, the volume correction height added to the topography elevation is given by
        \begin{equation}
            h_{i, \text{corr}} = \frac{f_{ij}} {\sum_{i=k}^{R} a_{\text{cell}}} \Delta V,
        \end{equation}
        where $f_{i,j}$ is the fraction covered by the cell [see~\eqref{eq:frac}], and $\sum_{i=k}^{R} a_{\text{cell}}$ is the total covered area of the intersecting cells.
        
        \section{Computation of averaged quantities}
        
        The average mean thickness $\overline{\langle h^{+} \rangle}$ is given by averaging the mean thickness within cells with non-zero thickness
        \begin{equation}
            \overline{\langle h^{+} \rangle} = \frac{1}{N} \sum_{k=1}^N \frac{1}{M^{+}_k} \sum_{ij} h^k_{ij},
            \label{eq:avg_thickness_mean}
        \end{equation}
        where $N$ is the number of runs (60), $M^{+}_k$ is the number of non-"empty" cells (non-zero thickness) in the $k$-th run [see also \eqref{eq:m_nonzero_cells}] and $h^k_{ij}$ is the thickness of the lava deposit in the cell $(i,j)$ in the $k$-th run.
        
        The average maximum thickness is the maximum thickness averaged over all runs
        \begin{equation}
            \overline{ \max(h) } = \frac{1}{N} \sum_{k=1}^N \max_{ij} \left(h^{k}_{ij}\right).
            \label{eq:avg_max_thickness}
        \end{equation}
        
        The average inundated area $\overline{A}$ is given by the cumulative area of the average number of cells with non-zero lava thickness,
        \begin{equation}
            \overline{A} = \frac{a_\text{cell}}{N} \sum_k M^+_k,
            \label{eq:avg_area}
        \end{equation}
        where $a_\text{cell}$, is the area of a single grid cell and $M^+_k$ is the number of cells with non-zero thickness [see also \eqref{eq:m_nonzero_cells}].
        
        For all of the aforementioned quantities the standard deviation $\sigma_{\overline{x}}$ is found from
        \begin{equation}
            \sigma_{\overline{x}} = \sqrt{ \frac{1}{N(N-1)}\sum_{k=1}^{N}\left(x_k - \overline{x}\right)^2},
            \label{eq:std_total}
        \end{equation}
        where $x_k$, with $x \in \{ \langle h \rangle, \max(h), A \}$, is the value of the quantity in the $k$-th run and $\overline{x}$ is the average over all runs.
        
        The average thickness in the cell $(i,j)$ is computed as
        \begin{equation}
            \overline{h_{ij}} = \frac{1}{N^{+}_{ij}} \sum_{k=1}^N h^k_{ij},
            \label{eq:avg_thickness_cell}
        \end{equation}
        where $N^{+}_{ij}$ is the total number of runs in which the cell was touched by lava, see \eqref{eq:n_nonzero_runs}, and $h^k_{ij}$ is the thickness of the lava deposited in the cell in the $k$-th run.
        Further, we compute the standard deviation of the thickness per cell as
        \begin{equation}
            \sigma_{\overline{h_{ij}}} = \sqrt{\frac{1}{N^{+}_{ij}(N^{+}_{ij} - 1)}  \sum_{k=1}^N \left(h^k_{ij} - \overline{h_{ij}}\right)^2}.
            \label{eq:std_thickness_cell}
        \end{equation}
        Defining the indicator function $c^k_{ij}$ as
        \begin{equation}
            c^k_{ij} = 
            \left\{\begin{matrix}
                0 & \text{if} & h^k_{ij} \leq 0\\
                1 & \text{if} & h^k_{ij} > 0
            \end{matrix}\right.,
            \label{eq:indicator}
        \end{equation}
        allows us to obtain $N^{+}_{ij}$ as
        \begin{equation}
            N^{+}_{ij} = \sum_{k=1}^{N} c^{k}_{ij},
            \label{eq:n_nonzero_runs}
        \end{equation}
        and $M^{+}_k$ as
        \begin{equation}
            M^{+}_k = \sum_{ij} c^{k}_{ij}.
            \label{eq:m_nonzero_cells}
        \end{equation}
        The average total volume can be computed as 
        \begin{equation}
            \overline{V} = \frac{a_\text{cell}}{N} \sum_k \sum_{ij} h^k_{ij}.
        \end{equation}
        
        We define a dimensionless relative RMS error, $\Delta_{\text{rel}}$ (sometimes also called the normalized RMS error) as
        
        \begin{equation}\label{eq:rel_rms}
            \Delta_{\text{rel}} = \left[ \frac{1}{N_{ij}^*} \sum_{ij} \left( \frac{\overline{\langle f_{ij}\rangle}}{\overline{\langle f \rangle}} - \frac{\overline{\langle g_{ij} \rangle}}{\overline{\langle g \rangle}} \right)^2\right]^{1/2},
        \end{equation}
        
        where $\overline{\langle f_{ij}\rangle}$ and $\overline{\langle g_{ij}\rangle}$ are the average non-zero thickness or masked non-zero averaged thickness of Flowy and MrLavaLoba for a current cell, respectively [see~\eqref{eq:avg_thickness_mean}]. 
        The height (or masked height) averaged spatially and over all runs is denoted by $\overline{\langle f \rangle}$ and $\overline{\langle g \rangle}$ for Flowy and the MrLavaLoba code (see Table~\ref{tab:averaged_quantities} for values of $\overline{\langle h \rangle}$ used).
        $N_{ij}^*$ is the total number of cells excluding the cells where both $\overline{\langle f_{ij}\rangle}$ and $\overline{\langle g_{ij}\rangle}$ are zero.

        \section{Coverage index}
        Given two thickness profiles $f_{ij}$ and $g_{ij}$, defined on the same grid, the coverage convergence index, $\mu$, quantifies the agreement between two inundated areas (without explicit reference to the thickness), such that a maximum value of $1$ indicates perfect agreement.
        It is defined as
        \begin{equation}
            \mu = \frac{ \left| \mathcal{A}_f \cap \mathcal{A}_g \right|}{ \left| \mathcal{A}_f \cup \mathcal{A}_g \right| },
            \label{eq:coverage_index}
        \end{equation}
        where $\mathcal{A}_{d}$ denotes the set of inundated cells
        \begin{equation}
            \mathcal{A}_{d} = \Big\{ (i,j)\;|\; d_{ij} > 0 \Big\} \text{ with  } d \in \{f, g\}.
        \end{equation}
        
        To compute the coverage index $\mu(N_\text{thresh})$, filtered by the sampling count with a threshold of $N_\text{thresh}$, Eq.~\eqref{eq:coverage_index} is used with a modified set of inundated cells
        \begin{equation}
            \mathcal{A}_{d}(N_\text{thresh}) = \Big\{ (i,j)\;|\; N^{+}_{ij} > N_\text{thresh} \Big\} \text{ with  } d \in \{f, g\}.
            \label{eq:filtered_convergence_index}
        \end{equation}
        
        \section{Comparison of thickness with masking}\label{appendix:masked_thickness}
        
        \begin{figure*}[thb]
            \centering
            \includegraphics[width=\linewidth]{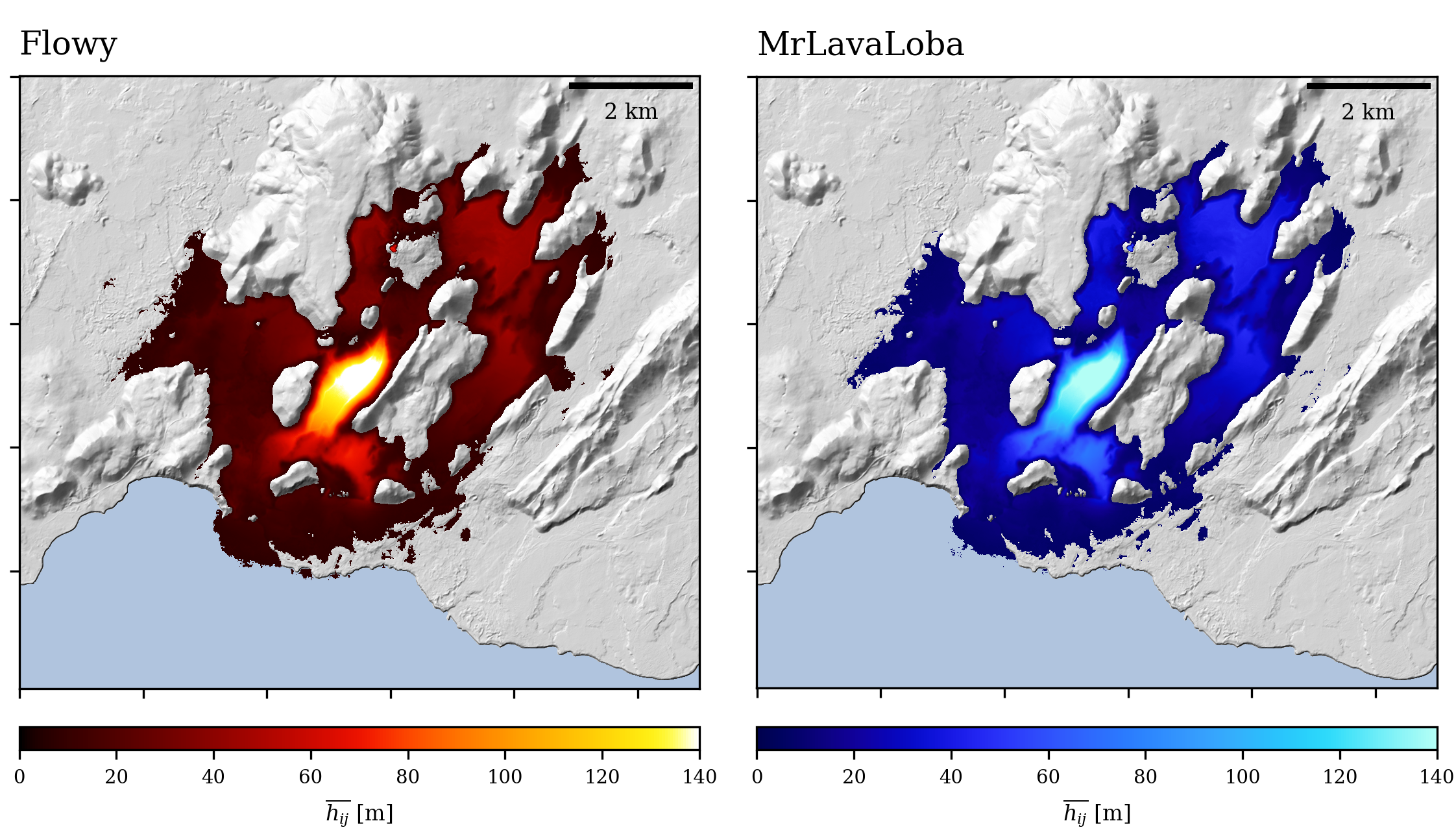}
            \caption{Comparison of the average cell resolved volume masked thickness (threshold=0.96) $\overline{h_{ij}}$, for the Flowy code (left, red) and the MrLavaLoba code (right, blue).
            The shade indicates the thickness of the lava deposit, with darker shades meaning thicker deposits.}
            \label{fig:comparison_thickness_masked}
        \end{figure*}

        \begin{figure}[t]
            \centering
            \includegraphics[width=\linewidth]{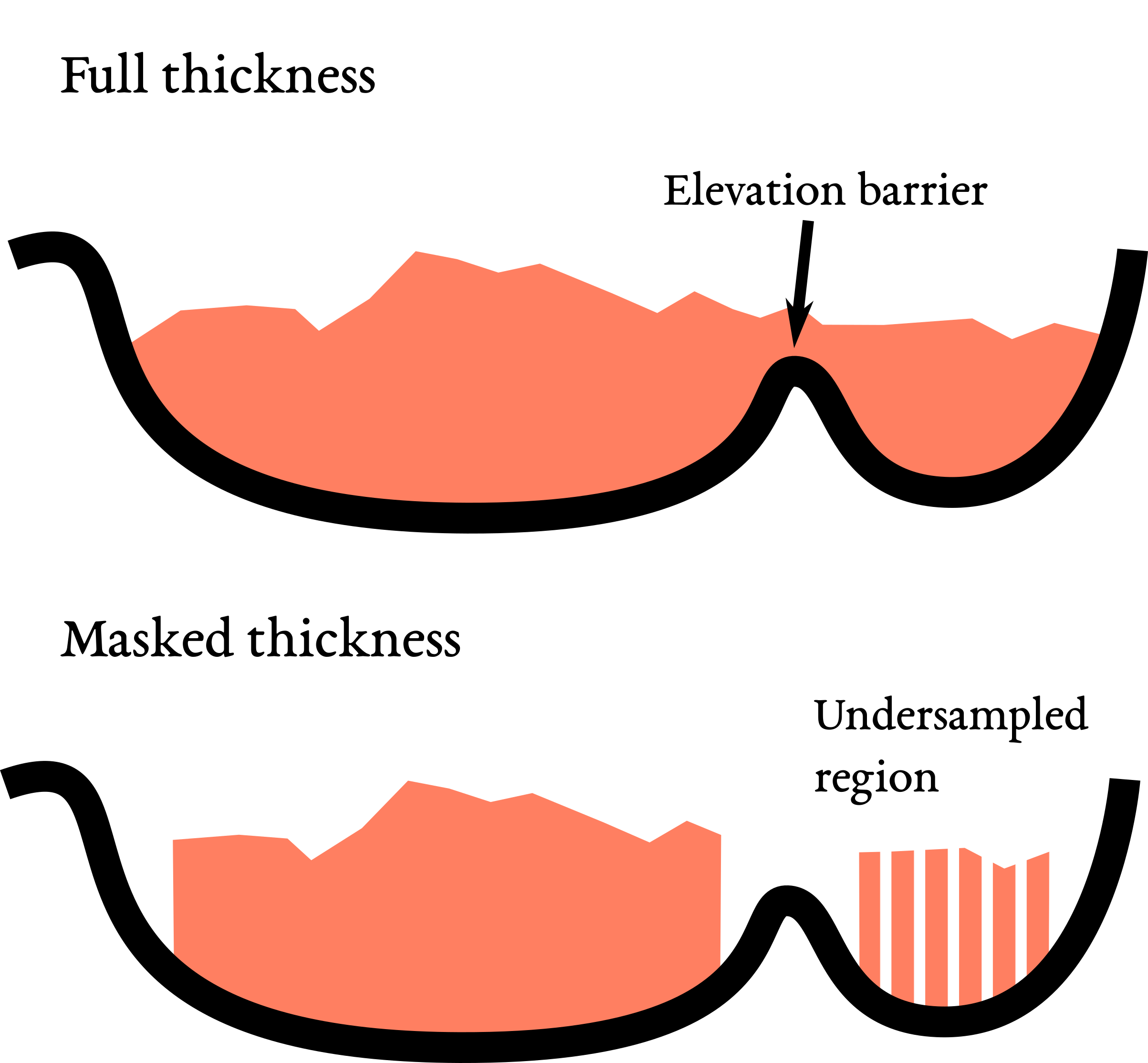}
            \caption{Schematic of a scenario with an elevation barrier separating a well-sampled area from a poorly sampled basin. 
            In most simulations, the lava is unable to surmount the barrier; however the barrier may be surmounted in a few simulations.
            In one such simulation, the undersampled basin is filled up.
            The masked thickness does not reflect this, and instead creates two discontinuous regions.
            The undersampled masked thickness is striped.}
        \label{fig:elevationBarrier}
        \end{figure}
        
        Figure~\ref{fig:comparison_thickness_masked} shows a comparison of the masked thickness, with masking threshold $0.96$, averaged over $60$ runs of Flowy and MrLavaLoba.
        
        The volume masking was proposed by Vitturi \emph{et al.} \cite{demichielivitturiMrLavaLobaNewProbabilistic2018} and removes all cells with a thickness below a certain threshold $t_\text{thresh}$.
        The threshold is calibrated such that the removed volume accounts for a certain fraction of the total volume.
        This fraction is determined by the masking threshold, such that a threshold of e.g. $0.96$ (like we used here) leads to 4 \% of the total volume being removed.
        
        In our case study in the main text, the relative RMS error for masked thicknesses was smaller than that of the unmasked thicknesses.
        
        However, the agreement is still not perfect.
        It should be kept in mind that this is expected, since the masking is not based on the sampling quality of the local thickness.
        To illustrate the difference, consider a situation in which a basin is far away from the vent and that filling up this basin is unlikely, because of an elevation barrier, as schematically shown in Figure~\ref{fig:elevationBarrier}.
        In most simulations, the lava tends to flow around the barrier; however, in some simulations, the barrier is overcome and lava fills the basin.
        Therefore, the basin is an undersampled region.
        The sampling count would correctly determine this basin to be a poorly sampled region.
        However, masking the thickness would just create two disconnected regions of lava, as depicted in Figure~\ref{fig:elevationBarrier}.
        
        To improve the agreement of the inundated areas, a criterion based on the sampling count, as shown in the main text, is better suited.
        
        \section{Tie-breaking selection rule for preliminary budding point} \label{appendix:tiebreaker}
        
        In the highly unlikely event that multiple perimeter points have exactly the same interpolated elevation, the tie-breaking rule selects the first point $(x,y) = (\sin \phi, \cos \phi)$ encountered when scanning the azimuthal angle $\phi \in [0, 2\pi)$. While this introduces a minor direction bias on a perfectly flat topography, the bias is quickly mitigated by the deposits from subsequent lobes (See Fig. S1). This effect is negligible in the context of realistic topographic conditions.
        
        \section{Probabilistic hazard maps}
        \label{sec:probabilistic_hazard}
        
        In Section~\ref{sec:hazard_map}, a qualitative measure of the hazard, related to the number of cumulative descendents in a flow, was defined. Such qualitative calculations can be used to create hazard maps in the form of different zones or even circular regions~\cite{loughlin2015global}. However, hazard modelling can be leveraged to create probabilistic hazard maps, which depict the inundation probability of an area~\cite{loughlin2015global}. In the context of Flowy, the probabilistic hazard can be defined as the probability that a cell in the topography is touched by lava. A probabilistic hazard map can, therefore, be obtained from the sampling count $N_{ij}^{+}$ (Eq.~\ref{eq:n_nonzero_runs}), normalized by $N$, the number of simulation runs.
        
        \clearpage
        \bibliographystyle{elsarticle-num}
        \bibliography{flowyPaper.bib}

\begin{thebibliography}{10}
\expandafter\ifx\csname url\endcsname\relax
  \def\url#1{\texttt{#1}}\fi
\expandafter\ifx\csname urlprefix\endcsname\relax\def\urlprefix{URL }\fi
\expandafter\ifx\csname href\endcsname\relax
  \def\href#1#2{#2} \def\path#1{#1}\fi

\bibitem{demichielivitturiMrLavaLobaNewProbabilistic2018}
M.~{de' Michieli Vitturi}, S.~Tarquini, {{MrLavaLoba}}: {{A}} new probabilistic
  model for the simulation of lava flows as a {\emph{settling}} process,
  Journal of Volcanology and Geothermal Research 349 (2018) 323--334.
\newblock \href {https://doi.org/10.1016/j.jvolgeores.2017.11.016}
  {\path{doi:10.1016/j.jvolgeores.2017.11.016}}.

\bibitem{lava_loba_github}
{{MrLavaLoba}}, \url{https://github.com/demichie/MrLavaLoba}.

\bibitem{flowy_github}
Flowy, \url{https://github.com/flowy-code/flowy}.

\bibitem{richardson2019multi}
P.~Richardson, L.~Karlstrom, The multi-scale influence of topography on lava
  flow morphology, Bulletin of Volcanology 81 (2019) 1--17.

\bibitem{kelfoun2016volcflow}
K.~Kelfoun, S.~V. Vargas, {VolcFlow} capabilities and potential development for
  the simulation of lava flows, Geological Society, London, Special
  Publications 426~(1) (2016) 337--343.

\bibitem{vicari2007modeling}
A.~Vicari, H.~Alexis, C.~Del~Negro, M.~Coltelli, M.~Marsella, C.~Proietti,
  Modeling of the 2001 lava flow at {E}tna volcano by a cellular automata
  approach, Environmental Modelling \& Software 22~(10) (2007) 1465--1471.

\bibitem{cappelloModelingGeophysicalFlows2022}
A.~Cappello, G.~Bilotta, G.~Ganci, Modeling of geophysical flows through
  {{GPUFLOW}}, Applied Sciences 12~(9) (2022) 4395.
\newblock \href {https://doi.org/10.3390/app12094395}
  {\path{doi:10.3390/app12094395}}.

\bibitem{harrisFLOWGOKinematicThermorheological2001}
A.~Harris, R.~S., {{FLOWGO}}: A kinematic thermo-rheological model for lava
  flowing in a channel, Bulletin of Volcanology 63~(1) (2001) 20--44.
\newblock \href {https://doi.org/10.1007/s004450000120}
  {\path{doi:10.1007/s004450000120}}.

\bibitem{harrisSimulatingThermorheologicalEvolution2016}
A.~J.~L. Harris, M.~Rh{\'e}ty, L.~Gurioli, N.~Villeneuve, R.~Paris, Simulating
  the thermorheological evolution of channel-contained lava: {{FLOWGO}} and its
  implementation in {{EXCEL}}, Geological Society, London, Special Publications
  426~(1) (2016) 313--336.
\newblock \href {https://doi.org/10.1144/SP426.9} {\path{doi:10.1144/SP426.9}}.

\bibitem{chevrelPyFLOWGOOpensourcePlatform2018}
M.~O. Chevrel, J.~Labroqu{\`e}re, A.~J.~L. Harris, S.~K. Rowland, {{PyFLOWGO}}:
  {{An}} open-source platform for simulation of channelized lava
  thermo-rheological properties, Computers \& Geosciences 111 (2018) 167--180.
\newblock \href {https://doi.org/10.1016/j.cageo.2017.11.009}
  {\path{doi:10.1016/j.cageo.2017.11.009}}.

\bibitem{cordonnier2016benchmarking}
B.~Cordonnier, E.~Lev, F.~Garel, Benchmarking lava-flow models, Geological
  Society, London, Special Publications 426~(1) (2016) 425--445.

\bibitem{dietterichBenchmarkingComputationalFluid2017}
H.~R. Dietterich, E.~Lev, J.~Chen, J.~A. Richardson, K.~V. Cashman,
  Benchmarking computational fluid dynamics models of lava flow simulation for
  hazard assessment, forecasting, and risk management, Journal of Applied
  Volcanology 6~(1) (2017) 9.
\newblock \href {https://doi.org/10.1186/s13617-017-0061-x}
  {\path{doi:10.1186/s13617-017-0061-x}}.

\bibitem{favalliForecastingLavaFlow2005}
M.~Favalli, M.~T. Pareschi, A.~Neri, I.~Isola, Forecasting lava flow paths by a
  stochastic approach, Geophysical Research Letters 32~(3) (2005).
\newblock \href {https://doi.org/10.1029/2004GL021718}
  {\path{doi:10.1029/2004GL021718}}.

\bibitem{mossouxQLAVHAFlexibleGIS2016}
S.~Mossoux, M.~Saey, S.~Bartolini, S.~Poppe, F.~Canters, M.~Kervyn,
  Q-{{LAVHA}}: {{A}} flexible {{GIS}} plugin to simulate lava flows, Computers
  \& Geosciences 97 (2016) 98--109.
\newblock \href {https://doi.org/10.1016/j.cageo.2016.09.003}
  {\path{doi:10.1016/j.cageo.2016.09.003}}.

\bibitem{tarquiniModelingLavaFlow2018}
S.~Tarquini, M.~de' Michieli~Vitturi, E.~Jensen, G.~Pedersen, S.~Barsotti,
  D.~Coppola, M.~Pfeffer, Modeling lava flow propagation over a flat landscape
  by using {{MrLavaLoba}}: The case of the 2014--2015 eruption at
  {{Holuhraun}}, {{Iceland}}, Annals of Geophysics 61~(Vol 61 (2018)) (2018) 5.
\newblock \href {https://doi.org/10.4401/ag-7812} {\path{doi:10.4401/ag-7812}}.

\bibitem{tarquiniAssessingImpactLava2020}
S.~Tarquini, M.~Favalli, M.~Pfeffer, M.~De' Michieli~Vitturi, S.~Barsotti,
  G.~Pedersen, B.~A. {\'O}lad{\'o}ttir, E.~H. Jensen, Assessing the impact of
  lava flows during the unrest of {{Svartsengi}} volcano in the {{Reykjanes}}
  peninsula, {{Iceland}}, in: Proceedings of the {{Geomorphometry}} 2020
  {{Conference}}, IRPI CNR, 2020.

\bibitem{barsottiEruptionFagradalsfjall20212023}
S.~Barsotti, M.~M. Parks, M.~A. Pfeffer, B.~A. {\'O}lad{\'o}ttir, T.~Barnie,
  M.~M. Titos, K.~J{\'o}nsd{\'o}ttir, G.~B.~M. Pedersen, {\'A}.~R.
  Hjartard{\'o}ttir, G.~Stefansd{\'o}ttir, T.~Johannsson, {\TH}.~Arason, M.~T.
  Gudmundsson, B.~Oddsson, R.~H. {\TH}rastarson, B.~G. {\'O}feigsson,
  K.~Vogfj{\"o}rd, H.~Geirsson, T.~Hj{\"o}rvar, S.~{von L{\"o}wis}, G.~N.
  Petersen, E.~M. Sigur{\dh}sson, The eruption in {{Fagradalsfjall}} (2021,
  {{Iceland}}): How the operational monitoring and the volcanic hazard
  assessment contributed to its safe access, Natural Hazards 116~(3) (2023)
  3063--3092.
\newblock \href {https://doi.org/10.1007/s11069-022-05798-7}
  {\path{doi:10.1007/s11069-022-05798-7}}.

\bibitem{pedersenLavaFlowHazard2023}
G.~B.~M. Pedersen, M.~A. Pfeffer, S.~Barsotti, S.~Tarquini, M.~{de'Michieli
  Vitturi}, B.~A. {\'O}lad{\'o}ttir, R.~H. {\TH}rastarson, Lava flow hazard
  modeling during the 2021 {{Fagradalsfjall}} eruption, {{Iceland}}:
  Applications of {{MrLavaLoba}}, Natural Hazards and Earth System Sciences
  23~(9) (2023) 3147--3168.
\newblock \href {https://doi.org/10.5194/nhess-23-3147-2023}
  {\path{doi:10.5194/nhess-23-3147-2023}}.

\bibitem{UnrestGrindavik}
Land continues to rise at {{Svartsengi}} {\textbar} {{News}},
  https://en.vedur.is/about-imo/news/volcanic-unrest-grindavik.

\bibitem{glazeSimulationInflatedPahoehoe2013}
L.~S. Glaze, S.~M. Baloga, Simulation of inflated pahoehoe lava flows, Journal
  of Volcanology and Geothermal Research 255 (2013) 108--123.
\newblock \href {https://doi.org/10.1016/j.jvolgeores.2013.01.018}
  {\path{doi:10.1016/j.jvolgeores.2013.01.018}}.

\bibitem{rewNetCDFInterfaceScientific1990}
R.~Rew, G.~Davis, {{NetCDF}}: An interface for scientific data access, IEEE
  Computer Graphics and Applications 10~(4) (1990) 76--82.
\newblock \href {https://doi.org/10.1109/38.56302}
  {\path{doi:10.1109/38.56302}}.

\bibitem{loughlin2015global}
S.~C. Loughlin, R.~S.~J. Sparks, S.~K. Brown, S.~F. Jenkins, C.~Vye-Brown,
  Global volcanic hazards and risk, Cambridge University Press, 2015.

\bibitem{pybind11}
W.~Jakob, J.~Rhinelander, D.~Moldovan, pybind11 -- {{Seamless}} operability
  between {{C++11}} and {{Python}}, \url{https://github.com/pybind/pybind11}
  (2017).

\bibitem{pyflowy_github}
{{PyFlowy}}, \url{https://github.com/flowy-code/pyflowy}.

\bibitem{kovesiGoodColourMaps2015}
P.~Kovesi, Good colour maps: How to design them (Sep. 2015).
\newblock \href {http://arxiv.org/abs/1509.03700} {\path{arXiv:1509.03700}},
  \href {https://doi.org/10.48550/arXiv.1509.03700}
  {\path{doi:10.48550/arXiv.1509.03700}}.

\bibitem{dem_2024}
\href{https://doi.org/10.5281/zenodo.11453396}{{DEM} of {I}celand in {E}sri
  {ASCII} format} (Jun. 2024).
\newblock \href {https://doi.org/10.5281/zenodo.11453396}
  {\path{doi:10.5281/zenodo.11453396}}.
\newline\urlprefix\url{https://doi.org/10.5281/zenodo.11453396}

\bibitem{delpImageCompressionUsing1979}
E.~Delp, O.~Mitchell, Image compression using block truncation coding, IEEE
  Transactions on Communications 27~(9) (1979) 1335--1342.
\newblock \href {https://doi.org/10.1109/TCOM.1979.1094560}
  {\path{doi:10.1109/TCOM.1979.1094560}}.

\bibitem{habibiHybridCodingPictorial1974}
A.~Habibi, Hybrid coding of pictorial data, IEEE Transactions on Communications
  22~(5) (1974) 614--624.
\newblock \href {https://doi.org/10.1109/TCOM.1974.1092258}
  {\path{doi:10.1109/TCOM.1974.1092258}}.

\bibitem{jainImageDataCompression1981}
A.~Jain, Image data compression: A review, Proceedings of the IEEE 69~(3)
  (1981) 349--389.
\newblock \href {https://doi.org/10.1109/PROC.1981.11971}
  {\path{doi:10.1109/PROC.1981.11971}}.

\bibitem{lindeAlgorithmVectorQuantizer1980}
Y.~Linde, A.~Buzo, R.~Gray, An algorithm for vector quantizer design, IEEE
  Transactions on Communications 28~(1) (1980) 84--95.
\newblock \href {https://doi.org/10.1109/TCOM.1980.1094577}
  {\path{doi:10.1109/TCOM.1980.1094577}}.

\bibitem{wilsonNewMetricGreyScale1997}
D.~L. Wilson, A.~J. Baddeley, R.~A. Owens, A new metric for grey-scale image
  comparison, International Journal of Computer Vision 24~(1) (1997) 5--17.
\newblock \href {https://doi.org/10.1023/A:1007978107063}
  {\path{doi:10.1023/A:1007978107063}}.

\bibitem{QGIS_software}
{QGIS Development Team}, \href{https://www.qgis.org}{{{QGIS}} {{Geographic}}
  {{Information}} {{System}}} (2024).
\newline\urlprefix\url{https://www.qgis.org}

\end{thebibliography}


\begin{thebibliography}{1}
\expandafter\ifx\csname url\endcsname\relax
  \def\url#1{\texttt{#1}}\fi
\expandafter\ifx\csname urlprefix\endcsname\relax\def\urlprefix{URL }\fi
\expandafter\ifx\csname href\endcsname\relax
  \def\href#1#2{#2} \def\path#1{#1}\fi

\bibitem{MrLavaLobaNewProbabilistic2018}
M.~{de' Michieli Vitturi}, S.~Tarquini, {{MrLavaLoba}}: {{A}} new probabilistic
  model for the simulation of lava flows as a {\emph{settling}} process,
  Journal of Volcanology and Geothermal Research 349 (2018) 323--334.
\newblock \href {https://doi.org/10.1016/j.jvolgeores.2017.11.016}
  {\path{doi:10.1016/j.jvolgeores.2017.11.016}}.

\bibitem{islandsdem}
Islandsdemv0, https://atlas.lmi.is/dem.

\bibitem{pedersenLavaFlowHazard20232}
G.~B.~M. Pedersen, M.~A. Pfeffer, S.~Barsotti, S.~Tarquini, M.~{de'Michieli
  Vitturi}, B.~A. {\'O}lad{\'o}ttir, R.~H. {\TH}rastarson, Lava flow hazard
  modeling during the 2021 {{Fagradalsfjall}} eruption, {{Iceland}}:
  Applications of {{MrLavaLoba}}, Natural Hazards and Earth System Sciences
  23~(9) (2023) 3147--3168.
\newblock \href {https://doi.org/10.5194/nhess-23-3147-2023}
  {\path{doi:10.5194/nhess-23-3147-2023}}.

\end{thebibliography}


\newpage
\setcounter{section}{0}
\setcounter{page}{1}
\setcounter{figure}{0}
\setcounter{equation}{0}
\renewcommand{\thesection}{S\arabic{section}}
\renewcommand{\thepage}{s\arabic{page}}
\renewcommand{\thetable}{S\arabic{table}}
\renewcommand{\thefigure}{S\arabic{figure}}

\setcounter{affn}{0}
\resetTitleCounters

\makeatletter
\let\@title\@empty
\makeatother

\title{Supplemental Material for Flowy: High performance probabilistic lava emplacement prediction}

\makeatletter
\renewenvironment{abstract}{\global\setbox\absbox=\vbox\bgroup
  \hsize=\textwidth\def\baselinestretch{1}%
  \noindent\unskip\textbf{Contents}
 \par\medskip\noindent\unskip}
 {\egroup}

\def\ps@pprintTitle{%
     \let\@oddhead\@empty
     \let\@evenhead\@empty
     \def\@oddfoot{\footnotesize\itshape
        Supplementary Data for \ifx\@journal\@empty Elsevier
       \else\@journal\fi\hfill\today}%
     \let\@evenfoot\@oddfoot}
\makeatother

\startlist{toc}
\begin{abstract}
\vspace{-48pt}
\printlist{toc}{}{\section*{}}
\end{abstract}
\maketitle
\section*{}
\parindent0pt
\section{Method details of Vitturi \textit{et al.} and parameters}
\label{sec:supple_method_details}

A brief overview of the MrLavaLoba method of lava emplacement, proposed by Vitturi \textit{et al.} ~\citesupple{MrLavaLobaNewProbabilistic2018}, was provided in Section 2 and Figure 2(a) in the main text. Here, we describe some user-defined input parameters and how the method can be modified via these parameters. However, this is not exhaustive; please refer to the class \texttt{Flowy::Config::InputParams} for more details. 
To recapitulate, each simulation consists of $N_{\text{flows}}$ (parameter \texttt{n\_flows}) inter-dependent flows, and each flow is built up of $N_{\text{lobes}}$ lobes (input parameter \texttt{n\_lobes}). 
First, we list a few parameters that affect all lobes (initial lobes and budding lobes) in the flow:

\begin{itemize}
    \item \texttt{aspect\_ratio\_coeff}: Determines how the aspect ratio of the lobe is scaled with the slope.
    \item \texttt{max\_aspect\_ratio}: This is the maximum aspect ratio allowed for the lobes
    \item \texttt{total\_volume}: This is the total erupted volume and must be provided. 
    \item \texttt{lobe\_area} and \texttt{avg\_lobe\_thickness}: Either the lobe area or the average lobe thickness must be kept fixed, and the other dimension can be determined using the erupted volume. Typically, the \texttt{lobe\_area} is defined by the user (usually scaled with the resolution of the DEM), and the average thickness of the lobes is calculated. 
    \item \texttt{thickness\_ratio}: The thickness of the lobes can progressively increase or decrease in every flow, depending on the value of the \texttt{thickness\_ratio}. Note that the average thickness of all lobes is kept constant. Figure 2(b) and Figure 2(c) show how lobes increase in thickness as the flow progresses, for a \texttt{thickness\_ratio} value of $2$.
\end{itemize}

Parameters associated with various steps in the method are mentioned below (note that not all steps are listed here): 

\begin{enumerate}
    \item \emph{ Initial lobe placement}: The parameter \texttt{vent\_flag} determines how the center of the initial lobe will be determined. For instance, the lobe center can be set to a user-defined vent location. Multiple vents may be active, which can also be tuned via the \texttt{vent\_flag}. More details about how \texttt{vent\_flag} dictates the initial lobe position are provided in Step~\ref{step:vent_flag} in Section~\ref{sec:misc_formulas}.  
    \item \emph{Lobe propagation}
        \begin{itemize}
            \item The parent lobe is selected from all predecessor lobes, according to rules influenced by the parameter \texttt{lobe\_exponent} (see Step~\ref{step:parent_sel_rules} in Section~\ref{sec:misc_formulas} for details). 
            \item The initial budding point can then be optionally perturbed, to add some randomness to the lobe propagation direction. The perturbation depends on the previously calculated slope, as well as the parameter \texttt{max\_slope\_prob} (described in Step~\ref{step:angle_perturb_rules} in Section~\ref{sec:misc_formulas}). Figure 2(b) and Figure 2(c), in the main text, show how the value of \texttt{max\_slope\_prob} can affect the placement of lobes.
            \item The effect of inertia can also be added to the budding lobe, via the input parameter \\\texttt{inertial\_exponent} (see also Step~\ref{step:inertial_rules} in Section~\ref{sec:misc_formulas}). 
        \end{itemize}
        \item \emph{Budding lobe dimensions}: Once the budding point position has been finalized, the major axis direction is known. Therefore, the center of the budding lobe is calculated using the semi axes and the direction of the parent lobe center and the budding point. The overlap between the parent lobe and the budding lobe can be tuned via the input parameter \texttt{dist\_fact}.
\end{enumerate}

\section{Miscellaneous formulas used in the MrLavaLoba method} \label{sec:misc_formulas}
In this section, we detail the minutiae of miscellaneous formulas used in the lava emplacement method of Vitturi et al.~\citesupple{MrLavaLobaNewProbabilistic2018}, which have not been covered in the preceding section and in the main text. The formulas are not grouped in a particular order. 

\begin{enumerate}
    \item Determining \texttt{n\_lobes}: If the number of lobes per flow is not a fixed constant, it can optionally be drawn from a uniform probability distribution, within the range \texttt{min\_n\_lobes} and \texttt{max\_n\_lobes}. Alternatively, the number of lobes can also be set according to a function

    \begin{multline}
        \label{eqn:n_lobes_beta_law}
        \texttt{n\_lobes} = \text{nint}\Big(\texttt{min\_n\_lobes} + \\\frac{X}{2} (\texttt{max\_n\_lobes}-\texttt{min\_n\_lobes}) \Big),
    \end{multline}
    
    where nint is a function that rounds a number to the nearest integer, and

    \begin{equation}
        X(x) = \frac{x^{\alpha-1} (1-x)^{\beta-1}}{\Gamma(\alpha)},
        \label{eqn:beta_law}
    \end{equation}
    where $x=(1-\texttt{idx\_flow})/(\texttt{n\_flows}-1)$, and \texttt{idx\_flow} corresponds to the index of the current flow, which is an integer in the range $[0, \texttt{n\_flows})$. In ~\ref{eqn:beta_law}, $\alpha$ and $\beta$ are set by the user using the parameters \texttt{a\_beta} and \texttt{b\_beta}.
    Coincidentally, the right-hand side in~\ref{eqn:beta_law} is a beta probability density function (not a beta distribution). Please note that this means that \texttt{n\_lobes} can potentially be set to a value greater than \texttt{max\_n\_lobes}. In practice, \texttt{n\_lobes} is usually set to a constant value.
    \item \label{step:delta_lobe_thickness}Calculating \texttt{delta\_lobe\_thickness}: The lobe thickness, $t$, for each lobe typically varies for each lobe in a flow. This can be set according to the relation 

    \begin{equation}
        \label{eqn:delta_thickness}
        \texttt{delta\_lobe\_thickness} = \frac{2}{\texttt{n\_lobes}-1} \left( t_{\text{avg}} - t_{\text{min}} \right),
    \end{equation}

    where the average lobe thickness, $t_{\text{avg}}$, is given by

    \begin{equation}
        \label{eqn:avg_lobe_thickness}
        t_{\text{avg}} = \frac{2 \times \texttt{total\_volume}}{\texttt{n\_flows} \times a_{\text{lobe}} (\texttt{min\_n\_lobes}+\texttt{max\_n\_lobes})},
    \end{equation}

    and the minimum lobe thickness, $t_{\text{min}}$ is given by 

    \begin{equation}
        \label{eqn:min_lobe_thickness}
        t_{\text{min}} = \frac{2\times \texttt{thickness\_ratio}}{ t_{\text{avg}} (\texttt{thickness\_ratio} + 1) },
    \end{equation}

    where $a_{\text{lobe}}$ is the area of each lobe, and \texttt{delta\_lobe\_thickness}, \texttt{total\_volume}, \texttt{n\_flows}, \texttt{thickness\_ratio}, \texttt{min\_n\_lobes} and \texttt{max\_n\_lobes} are user-defined input parameters.
    
    \item \label{step:vent_flag} Getting initial lobe positions from \texttt{vent\_flag}: The positions of the initial lobes are calculated using certain rules, and, naturally do not originate from previously deposited lobes. How the lobes are positioned is controlled by the user-defined integer $\texttt{vent\_flag} \in [0,8]$. 
    
    Further, the method models effusive eruptions that can start from "fissures" as well as vents. Here, we define a fissure simply as a path, consisting of linear line segments, whose end points may be vents or other user-defined values. If such a fissure is defined, the initial lobes are placed on the fissure. We briefly describe the behaviour produced by different values of the \texttt{vent\_flag} in the following:
    \begin{itemize}
        \item $\texttt{vent\_flag}=0$: Initial lobes are positioned on vent coordinates. If there are multiple vents defined, each flow iteratively starts from the first, second, third vent, and so on.
        \item $\texttt{vent\_flag}=1$: Vents are randomly chosen, such that each vent has the same probability. Therefore, the vent is drawn from a uniform probability distribution. Each initial lobe is then situated on the selected vent.
        \item $\texttt{vent\_flag}=3$: In this case, the end points of each fissure are vent coordinates. The fissure chosen is selected from a uniform probability distribution of these defined fissures. Further, for each initial lobe, the point which defines the lobe center is selected from all the points that constitute the selected fissure, using a uniform probability distribution.  
        \item $\texttt{vent\_flag}=2$: Similar to $\texttt{vent\_flag}=3$, the end points of each fissure are vent coordinates. However, the difference is that the fissure chosen is weighted by the length of each fissure. Once a fissure has been chosen, the initial lobe center is selected from all the points that constitute the selected fissure, using a uniform probability distribution.
        \item $\texttt{vent\_flag}=5$: The fissure end points are defined such that the first point is a vent coordinate, and the end point is set from the user-defined \texttt{fissure\_end\_coordinates}. Thereafter, the fissure is chosen using a uniform probability distribution. Next, the initial lobe center is selected from all the points that constitute the selected fissure, using a uniform probability distribution.
        \item $\texttt{vent\_flag}=4$: Analogous to $\texttt{vent\_flag}=5$, the fissure end points are defined such that the first point is a vent coordinate, and the end point is set from the user-defined \texttt{fissure\_end\_coordinates}. Thereafter, the fissure is chosen such that the probability of selecting a fissure is weighted by its length. Next, the initial lobe center is selected from all the points that constitute the selected fissure, using a uniform probability distribution.
        \item $\texttt{vent\_flag}=6$: Analogous to $\texttt{vent\_flag}=3$, the fissure end points are vent coordinates. Each fissure is chosen such that the probability of selecting a fissure is determined using user-defined \texttt{fissure\_probabilities}. The initial lobe center is selected from all the points that constitute the selected fissure, using a uniform probability distribution.
        \item $\texttt{vent\_flag}=7$: Similar to $\texttt{vent\_flag}=5$, the fissure end points are defined such that the first point is a vent coordinate, and the end point is set from the user-defined \texttt{fissure\_end\_coordinates}. Each fissure is chosen such that the probability of selecting a fissure is determined using user-defined \texttt{fissure\_probabilities}. The initial lobe center is selected from all the points that constitute the selected fissure, using a uniform probability distribution.
        \item $\texttt{vent\_flag}=8$: In this case, a fissure is not defined. However, a vent is selected using the user-defined (and misleadingly named) \texttt{fissure\_probabilities}. This is accomplished by drawing the vent from a discrete probability distribution, weighted by the \texttt{fissure\_probabilities}. The initial lobe center is then set to the chosen vent coordinates. 
    \end{itemize}
    Note that using \texttt{vent\_flag} values of 0, 1 or 8 may lead to unphysical stacking of initial lobes over multiple flows, especially if there is only one vent. This is because the initial lobes are placed directly over vent coordinates in these cases (per flow).
    \item Determining the thickness of lobes: The thickness, $t$, of a lobe (initial or budding) is set according to 

    \begin{multline}
        \label{eqn:lobe_thickness}
        t = (1 -\texttt{thickening\_parameter}) \\\times (t_{\text{min}} + \texttt{idx\_lobe} \times \texttt{delta\_lobe\_thickness}),
    \end{multline}

    where \texttt{idx\_lobe} is the integral index of the current lobe such that $\texttt{idx\_lobe} \in [0, \texttt{n\_lobes})$, \\ and \texttt{delta\_lobe\_thickness} is a user-defined parameter. See Equations~\ref{eqn:delta_thickness},  ~\ref{eqn:min_lobe_thickness} for definitions of \texttt{delta\_lobe\_thickness} and $t_{\text{min}}$.

    \item \label{step:parent_sel_rules}Parent lobe selection: During the process of propagating lobes, for each budding lobe, a parent lobe must be selected out of all the previously created lobes. As mentioned in the preceding Section~\ref{sec:supple_method_details}, the parent lobe for a particular budding lobe is selected according to rules influenced by the user-defined parameter \texttt{lobe\_exponent}, which are described in the following.
    \begin{itemize}
        \item $\texttt{lobe\_exponent}>=1$: The parent lobe index is drawn from a uniform probability distribution.
        \item $\texttt{lobe\_exponent}<=0$: The parent lobe is always the last lobe created.
        \item $\texttt{lobe\_exponent} \in (0, 1)$: The parent lobe index, $i_{\text{parent}}$, is selected such that 

        \begin{equation}
            \label{eqn:parent_lobe_index}
            i_{\text{parent}} = i_{\text{descendent}} \times m
        \end{equation}

        where $i_{\text{descendent}}$ is the index of the budding lobe, and $m$ is drawn from the following distribution

        \begin{equation}
            \label{eqn:dist}
            P(x) = \frac{1}{\epsilon} x^{ \frac{1}{\epsilon} -1 },
        \end{equation}
    \end{itemize}

    where $\epsilon = \texttt{inertial\_exponent} \in (0,1)$.
    
    \item \label{step:angle_perturb_rules}Angle perturbation: As mentioned in the main text, a perturbation can be added to the lobe semi-major axis direction (for both initial and budding lobes). The extent of the angle perturbation is dictated by the value of the user-defined parameter \texttt{max\_slope\_prob}. We describe how the angle perturbation is calculated depending on the value of \texttt{max\_slope\_prob} in the following. 
    \begin{itemize}
        \item $\texttt{max\_slope\_prob}=1$: The angle perturbation is 0. Therefore, the azimuthal angle, $\phi$, remains the same. 
        \item $\texttt{max\_slope\_prob}=0$: The angle perturbation is drawn from a uniform probability distribution.
        \item $\texttt{max\_slope\_prob} \in (0,1)$: The angle perturbation is drawn from a truncated normal distribution, constrained to the interval $[-\pi,\pi]$. This truncated normal distribution has a mean of $0$. The standard deviation $\sigma$ of the normal distribution, which we truncate, is given by 

    \begin{equation}
        \label{eqn:angle_perturbation}
        \sigma = \frac{\pi}{180} \frac{(1-\texttt{max\_slope\_prob})(90-s')}{s' \times \texttt{max\_slope\_prob}},
    \end{equation}
    where the user-defined parameter \texttt{max\_slope\_prob} dictates the extent of the perturbation, and $s'$ is the local slope, in degrees. Note, however, that $\sigma$ is in radians, and, in fact, the azimuthal angle is also in radians. In future, it would be best to re-define \texttt{max\_slope\_prob} to reflect the fact that angles in Flowy are calculated in radians. 
    \end{itemize}

     \item \label{step:inertial_rules}Inertial contribution: This is touched upon in the main text. An inertial contribution can also, optionally, be added to the lobe semi-major axis direction of budding lobes (but not initial lobes, which do not have a parent). Let the current azimuthal angle of the budding lobe under consideration be $\phi$, and that of the parent lobe be $\phi_{\text{parent}}$. The user-defined parameter \texttt{inertial\_exponent} dictates the value of the inertial contribution added to $\phi$.
     \begin{itemize}
         \item $\texttt{inertial\_exponent}=0$: No inertial contribution is added to $\phi$. The value of $\phi$ is unchanged.
         \item $\texttt{inertial\_exponent}>0$: The new azimuthal angle, $\phi'$, is obtained from $\phi$ and $\phi_{\text{parent}}$ via the following relation

         \begin{equation}
             \label{eqn:inertial}
             \phi' = \arctan \left( \frac{ \cos\phi(1-\alpha_{\text{inertial}}) + \alpha_{\text{inertial}} \cos\phi_{\text{parent}} }{ \sin\phi(1-\alpha_{\text{inertial}}) + \alpha_{\text{inertial}} \sin\phi_{\text{parent}} } \right), 
         \end{equation}

         where $\alpha_{\text{inertial}}$ is given by

         \begin{equation}
             \label{eqn:alpha_inertial}
             \alpha_{\text{inertial}} = \left[ 1 - \left( \frac{2 \arctan(s)}{\pi} \right)^{ \eta } \right]^{ \frac{1}{\eta} },
         \end{equation}
         where $\eta = \texttt{inertial\_exponent}$, and $s$ is the local slope.
     \end{itemize}
     
     \item Determining semi-major and minor axes: The semi-major axes, $a$ and $b$, can be determined with the help of certain user-defined parameters and the local slope, as mentioned in Section~\ref{sec:supple_method_details}, given by 

     \begin{equation}
         \label{eqn:semi-major_axis}
         a = \left(\frac{ l a_{\text{lobe}}}{\pi} \right)^{\frac{1}{2}},
     \end{equation}

     and 

     \begin{equation}
         \label{eqn:semi-minor_axis}
         b = \left(\frac{ a_{\text{lobe}}}{l\pi}\right)^{ \frac{1}{2} },
     \end{equation}

     where $a_{\text{lobe}}$ is the lobe area and $l$ is the aspect ratio calculated according to 

     \begin{equation}
         \label{eqn:aspect_ratio}
         l = \min (\texttt{max\_aspect\_ratio}, 1+s \times \texttt{aspect\_ratio\_coeff}),
     \end{equation}

     where $s$ is the local slope, and \texttt{max\_aspect\_ratio} and \texttt{aspect\_ratio\_coeff} are user-defined parameters. 

     \item Lobe center of budding lobes: This is a moot point for initial lobes, since the lobe centers are determined first using rules defined by \texttt{vent\_flag} (see Step~\ref{step:vent_flag}). However, for budding lobes, the last step prior to rasterizing the budding lobes is to calculate the lobe center, $\vec{x}_{c}$, given by 

     \begin{equation}
         \label{eqn:budding_lobe_center}
         \vec{x}_{c} = \vec{x}_{b}' + a f \frac{(\vec{x}_{b}'-\vec{x}_{c\text{,parent}})}{||\vec{x}_{b}'-\vec{x}_{c\text{,parent}}||},
     \end{equation}

     where $\vec{x}_{b}'$ is the final budding point, $a$ is the budding lobe semi-major axis, $\vec{x}_{c\text{,parent}}$ is the center of the parent lobe, and $f$ is a user-defined parameter set by \texttt{dist\_fact}, whose value determines the degree of overlap between the parent lobe and descendant lobe. 
\end{enumerate}

\section{DEM used for eruption scenario starting from Geldingadalir valley}

The method of lava emplacement, implemented in Flowy, requires an input Digital Elevation Model (DEM) of the topography. 
As described in the main text, the input topography should be in the form of a two-dimensional grid. We have used a publicly available DEM of Iceland, called IslandsDEMv0~\citesupple{islandsdem}, which has a resolution of 2~m.
This topography was then downsampled to a resolution of 10~m.

\section{Input parameters for Geldingadalir valley case study}

Table \ref{tab:input} lists all used input parameters for the Geldingadalir valley case study, in the Reykjanes peninsula.
Certain parameters were derived from the work of Pedersen \textit{et al}~\citesupple{pedersenLavaFlowHazard20232}.

\begin{table}[t]
\centering
\caption{Input settings, used for the Geldingadalir valley case study. Since the MrLavaLoba code and Flowy support the same input file keywords, these are identical for both.}
\label{tab:input}
    \begin{tabular}{l r}
        \toprule
        \multicolumn{2}{c}{Input settings}\\
        \midrule
        vent\_flag             & 2\\
        x\_vent                & [339326, 339423]\\
        y\_vent                & [380202, 380364]\\
        east\_to\_vent         & 20880.0\\ 
        west\_to\_vent         & 20880.0\\ 
        south\_to\_vent        & 20880.0\\ 
        north\_to\_vent        & 20880.0\\ 
        hazard\_flag           & 1\\ 
        masking\_threshold     & 0.96\\ 
        n\_flows               & 180\\ 
        min\_n\_lobes          & 10000\\ 
        max\_n\_lobes          & 10000\\ 
        volume\_flag           & 1\\ 
        total\_volume          & 520000000\\ 
        fixed\_dimension\_flag & 1\\ 
        lobe\_area             & 1000.0\\ 
        thickness\_ratio       & 2\\ 
        topo\_mod\_flag        & 2\\ 
        n\_flows\_counter      & 1\\ 
        n\_lobes\_counter      & 1\\ 
        thickening\_parameter  & 0.06\\ 
        lobe\_exponent         & 0.015\\ 
        max\_slope\_prob       & 0.8\\ 
        inertial\_exponent     & 0.1\\ 
\midrule
        \multicolumn{2}{c}{``Advanced" input settings}\\
        \midrule
        restart\_files          & {[}{]} \\
        saveshape\_flag         & 0      \\
        saveraster\_flag        & 1      \\
        flag\_threshold         & 1      \\
        plot\_lobes\_flag       & 0      \\
        plot\_flow\_flag        & 0      \\
        a\_beta                 & 0.0    \\
        b\_beta                 & 0.0    \\
        force\_max\_length      & 0      \\
        max\_length             & 50     \\
        n\_init                 & 1      \\
        n\_check\_loop          & 0      \\
        start\_from\_dist\_flag & 0      \\
        dist\_fact              & 0.5    \\
        npoints                 & 30     \\
        aspect\_ratio\_coeff    & 2.0    \\
        max\_aspect\_ratio      & 2.5    \\
        shape\_name             & ``"     \\
        \bottomrule
    \end{tabular}
\end{table}

\section{Simulations on a perfectly flat topography}

\begin{figure}[h]
    \centering
    \includegraphics[width=\linewidth]{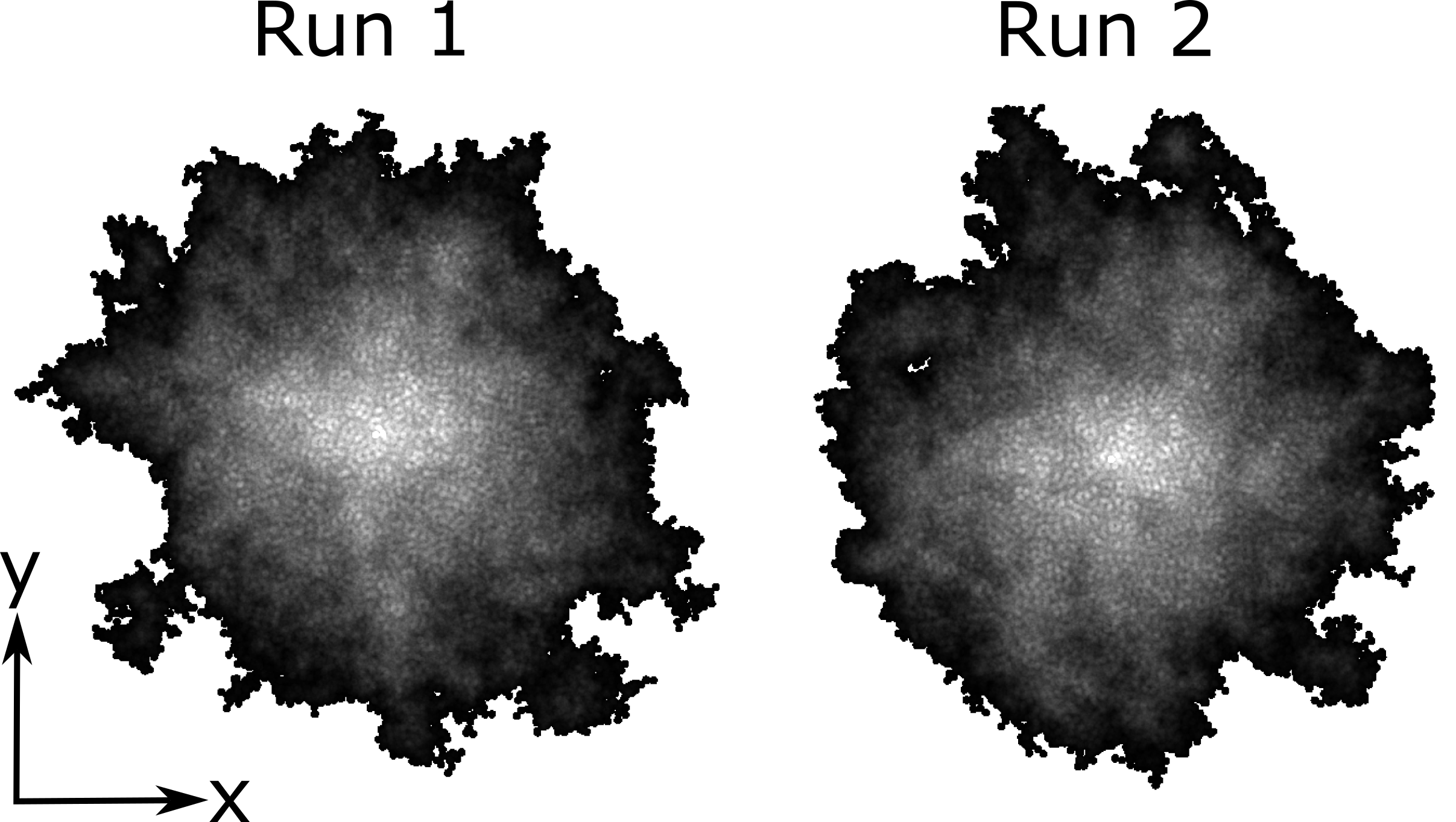}
    \caption{Height of lava deposits on a perfectly flat topography for two simulation runs in Flowy. Higher elevations are lighter in shade.}
\label{fig:spreadtest}
\end{figure}

Fig.~\ref{fig:spreadtest} shows the height of lava deposits on a perfectly flat toy topography after two simulation runs of Flowy. Such a toy topography is far from realistic, but is useful to illustrate how, without the effect of elevation features, lava should spread randomly. There appears to be no obvious bias in the y-dimension, despite the use of a tie-breaker rule for selecting preliminary budding points (Appendix J in the main text). 

\FloatBarrier
\bibliographystylesupple{elsarticle-num}
\bibliographysupple{supple_refs.bib}

\end{document}